\documentclass[showpacs,preprintnumbers,amsmath,amssymb,floatfix]{revtex4}

\usepackage{color}   
\usepackage{graphicx}
\usepackage{dcolumn}
\usepackage{bm}

\begin{document}
\def\bbox#1{\hbox{\boldmath${#1}$}}
\def\blambda{{\hbox{\boldmath $\lambda$}}}
\def\eeta{{\hbox{\boldmath $\eta$}}}
\def\bxi{{\hbox{\boldmath $\xi$}}}
\def\bzeta{{\hbox{\boldmath $\zeta$}}}

\title{ Quarkonia and Quark Drip Lines in Quark-Gluon Plasma}

\author{Cheuk-Yin Wong}

\affiliation{Physics Division, Oak Ridge National Laboratory, Oak Ridge, TN
37831}

\affiliation{Department of Physics, University of Tennessee, Knoxville, TN 
37996}
\date{\today}

\begin{abstract}

We extract the $Q$-$\bar Q$ potential by using the thermodynamic
quantities obtained in lattice gauge calculations. The potential is
tested and found to give spontaneous dissociation temperatures that
agree well with those from lattice gauge spectral function analysis.
Using such a $Q$-$\bar Q$ potential, we examine the quarkonium states
in a quark-gluon plasma and determine the `quark drip lines' which
separate the region of bound color-singlet $Q\bar Q$ states from the
unbound region.  The characteristics of the quark drip lines severely
limit the region of possible bound $Q\bar Q$ states with light quarks
to temperatures close to the phase transition temperature.  Bound
quarkonia with light quarks may exist very near the phase transition
temperature if their effective quark mass is of the order of 300-400
MeV and higher.

\end{abstract}

                                                                         
\maketitle

\section{Introduction}
                                                                            
The degree to which the constituents of a quark-gluon plasma (QGP) can
combine to form composite entities is an important property of the
plasma.  It has significant implications on the nature of the phase
transition, the quark-gluon plasma equation of state, the probability
of recombination of plasma constituents prior to the phase transition,
and the chemical yields of the observed bound hadrons.  The successes
of the recombination model
\cite{Phewp,Stawp,Brawp,Phowp,Hwa77,Hwa03,Gre03,Gre03a,Fri03,Fri05}
suggests that quark partons may form bound or quasi-bound states in
the quark-gluon plasma, at least at, or close to, the phase transition
temperature.  It is an important theoretical question as to the range
of temperatures in which these quarkonia may be bound or quasi-bound.
The successes of the thermal model
\cite{Phewp,Stawp,Brawp,Phowp,Bra95,Bra96,Cle99a,Bra96a,Cle99b,Cle99c,Bra02,Bra03,Bra04,Cle05}
for hadron yields also raises the important question whether hadrons
may become bound or quasi-bound in the quark-gluon plasma.  If they
are indeed bound in the quark-gluon plasma, the approach to chemical
equilibrium may commence in the quark-gluon plasma phase before the
phase transition and the boundaries of the quark-gluon plasma phase
and the hadron phase may overlap.

Recent spectral analyses of quarkonium correlators indicated that
$J/\psi$ may be stable up to 1.6$T_c$ where $T_c$ is the phase
transition temperature \cite{Asa03,Dat03}.  Subsequently, there has
been renewed interest in quarkonium states in quark-gluon plasma as
Zahed and Shuryak suggested that $Q \bar Q$ states with light quarks
may be bound up to a few $T_c$ \cite{Shu03}.  Quarkonium bound states
and instanton molecules in the quark-gluon plasma have been considered
by Brown, Lee, Rho, and Shuryak \cite{Bro04}.  As heavy quarkonia may
be used as a diagnostic tool \cite{Mat86}, there have been many recent
investigations on the stability of heavy quarkonia in the plasma
\cite{Won01a,Won05,Won05a,Won06a,Won06b,Dig01a,Won01a,Kac02,Kac03,Kac05,Alb05,Bla05,Son05,Pet05,Dig05}.

Previously, DeTar \cite{Det85}, Hansson, Lee, and Zahed \cite{Han88},
and Simonov \cite{Sim95} observed that the range of strong interaction
is not likely to change drastically across the phase transition and
suggested the possible existence of relatively narrow low-lying $Q\bar
Q$ states in the plasma.  On the other hand, Hatsuda and Kunihiro
\cite{Hat85} considered the persistence of soft modes in the plasma
which may manifest themselves as pion-like and sigma-like states.  The
use of baryon-strangeness correlation and charge fluctuation to study
the abundance of light quarkonium states in the plasma have been
suggested recently \cite{Koc05,Kar05}.

We would like to investigate the composite properties of the plasma
and to determine its `quark drip lines'.  We shall focus our attention
on color-singlet $Q$-$\bar Q$ states and the $Q$-$\bar Q$ quantities
in this paper refer to those of color-singlet $Q$-$\bar Q$ states
unless specified otherwise.  Here we follow Werner and Wheeler
\cite{Wer58} and use the term `drip line' to separate the region of
bound color-singlet $Q\bar Q$ states from the unbound region of
spontaneous quarkonium dissociation.  It should be emphasized that a
quarkonium can be dissociated by collision with constituent particles
to lead to the corresponding `particle-dissociation lines', which can
be interesting subjects for future investigation.

We would like to use the potential model to study the stability of
quarkonia, as the potential model can be used to evaluate many more
quantities than the lattice gauge spectral function analysis.  The
potential model lends itself to extrapolation into unknown regions of
quark masses and temperatures.  An important physical quantity in the
potential model is the $Q$-$\bar Q$ potential between the quark $Q$
and the antiquark $\bar Q$ at a separation $R$ at a temperature
$T$. Previous work in the potential model uses the color-singlet free
energy $F_1(R,T)$ \cite{Dig01a,Won01a,Bla05} or the color-singlet
internal energy $U_1(R,T)$ \cite{Kac02,Shu03,Alb05} obtained in
lattice gauge calculations as the $Q$-$\bar Q$ potential, without
rigorous theoretical justifications.  Here, the subscripts of
$U_1(R,T)$ and $F_1(R,T)$ refer to the color-singlet property of the
$Q$-$\bar Q$ system.  The internal energy $U_1(R,T)$ is significantly
deeper and spatially more extended than the free energy
$F_1(R,T)$. The degree of quarkonium binding will be significantly
different whether one uses the internal energy $U_1(R,T)$ or the free
energy $F_1(R,T)$ as the $Q$-$\bar Q$ potential.  Treating the
internal energy $U_1(R,T)$ as the $Q$-$\bar Q$ potential led Shuryak
and Zahed to suggest the possibility of color-singlet quarkonium
states with light quarks in the plasma \cite{Shu03}.  The conclusions
will be quite different if one uses the free energy $F_1(R,T)$ as the
$Q$-$\bar Q$ potential.

While $F_1(R,T)$ or $U_1(R,T)$ can both be used as the $Q$-$\bar Q$
potential at $T=0$ (at which $F_1(R,T)=U_1(R,T)$), the situation is
not so clear in a thermalized quark-gluon plasma.  It is important to
find out the meaning of these thermodynamical quantities calculated in
the finite-temperature lattice gauge theory so as to extract the
$Q$-$\bar Q$ potential.

If one constructs the Schr\" odinger equation for the $Q$ and $\bar Q$
in a thermal medium, the $Q$-$\bar Q$ potential in the Schr\" odinger
equation contains those interactions that act on $Q$ and $\bar Q$,
when the medium particles have re-arranged themselves
self-consistently.  On the other hand, the total internal energy
$U_1(R,T)$ contains not only these interactions that act on $Q$ and
$\bar Q$, but also the internal gluon energy $U_g(R,T)$ relative to
the gluon internal energy $U_{g0}$ in the absence of $Q$ and $\bar Q$,
as shown deductively in Ref.\ \cite{Won05} starting from the
definition of the free energy in lattice gauge theory in quenched QCD.
If the gluon internal energy $U_g(R,T)$ were independent of $R$, then
$U_1(R)$ could well be used as the $Q$-$\bar Q$ potential.  However,
in the grand canonical ensemble, $U_g(R,T)$ depends on $R$.  To get
the $Q$-$\bar Q$ potential, it is therefore necessary to subtract
$U_g(R,T)-U_{g0}$ from $U_1(R,T)$.  As the subtleties of these results
may not appear evident and the problem of non-perturbative QCD so
intrinsically complicated, a thorough understanding of an analogous,
but not identical, problem in QED is worth having.  Therefore, we
examine in detail the simple QED case of Debye screening of charges
$Q$ and $\bar Q$ in a massless charged medium in a grand canonical
ensemble, where the results can be readily obtained analytically.  We
would like to show that there is a relationship between the $Q$-$\bar
Q$ potential and the total internal energy when screening occurs: the
Debye screening $Q$-$\bar Q$ potential between two static opposite
charges in QED can be obtained from the total internal energy by
subtracting out the internal energy of the medium particles.

The results in the Debye screening case in QED support our previous
conclusion in Ref.\ \cite{Won05} that in the QCD lattice gauge
calculations in the grand canonical ensemble, it is necessary to
subtract out the $R$-dependent internal energy of the QGP from the
total internal energy in order to obtain the potential between $Q$ and
$\bar Q$ in the plasma.  Additional lattice gauge calculations may be
needed to evaluate the QGP internal energy in the presence of $Q$ and
$\bar Q$.  It is nonetheless useful at this stage to suggest
approximate ways to evaluate the QGP internal energy.  We proposed
earlier a method by making use of the equation of state of the
quark-gluon plasma obtained in an independent lattice gauge
calculation \cite{Won05}.  The equation of state provides a
relationship between the QGP internal energy and the QGP entropy
content.  As the QGP entropy content is the difference $U_1-F_1$, the
$Q$-$\bar Q$ potential can be represented as a linear combination of
$U_1$ and $F_1$, with coefficients depending on the quark-gluon plasma
equation of state.  The proposed potential was tested and found to
give spontaneous dissociation temperatures that agree well with those
from lattice gauge spectral function analysis in the quenched
approximation.

The comparison of the potential model results with those from spectral
analyses in the same quenched approximation is useful as a theoretical
test of the potential model.  However, in the quenched approximation,
the quark-gluon plasma is assumed to consist of gluons only and the
effects of dynamical quarks have not been included.  As dynamical
quarks provide additional screening, one wishes to know whether this
additional screening will modify the binding energies of quarkonia
significantly or not.  The presence of dynamical quarks also lowers
the phase transition temperature from 269 MeV for quenched QCD to 154
MeV for full QCD with three flavors \cite{Kar00}.  For these reasons,
it is necessary to include dynamical quarks to assess their effects on
the stability of quarkonia.  The knowledge of the single-particle
states using potentials extracted from lattice gauge calculations in
full QCD can then be used to examine the stability of both heavy and
light quarkonia and to determine the location of the quark drip lines.
We focus our attention mainly on heavy quarkonia for which a
non-relativistic treatment is a good description.  However, the
problem of the stability of quarkonia with light quarks is
intrinsically so complicated and the question of their stability up to
a temperature of few units of $T_c$ so important
\cite{Phewp}-\cite{Shu03} that even an approximate estimate using the
non-relativistic potential model is worth having.  The subject of
light quarkonia will be examined again, with the inclusion of the
relativistic effects as in recent works \cite{Cra04,Cra06}, in the
course of time.

The authors of Refs.\ \cite{Moc06,Moc06a} claim recently that
potential models cannot describe heavy quarkonia above $T_c$, as their
potential model correlators fail to reproduce lattice gauge
correlators {\it for all cases with all types of potentials. }  Such a
complete disagreement for all cases and all types of potentials
suggests that the lack of agreement may not be due to the potential
model (or models) themselves but to their method of evaluating the
meson correlators in the potential model.  We show recently that when
the contributions from the bound states and continuum states are
properly treated \cite{Won06b}, the potential model correlators
obtained with the proposed potential in Ref.\ \cite{Won05} are
consistent with lattice gauge correlators.

This paper is organized as follows.  In Section II, we describe the
puzzling behavior of the increase of the entropy of the QCD medium as
the separation between $Q$ and $\bar Q $ increases.  To understand
such a behavior, we introduce in Section III a simple model of Debye
screening in QED for which various thermodynamic quantities can be
readily calculated.  In Section VI, the variation of the number
density and entropy density of the medium particles for a $Q$-$\bar Q$
pair in Debye screening is shown to depend on the $Q$-$\bar Q$
separation $R$ when the second-order contributions are included.  In
Section V, we find similarly that the internal energy of the medium
particles in Debye screening also increases with $R$ when second-order
contributions are included.  In Section VI, we examine the Schr\"
odinger equation for the relative motion of $Q$ and $\bar Q$ and
identify the potential between $Q$ and $\bar Q$ in Debye screening.
We reach the conclusion that in order to obtain the Debye screening
potential between two static charges in the grand canonical ensemble,
it is necessary to subtract out the internal energy of the medium
particles from the total internal energy.  Returning to lattice gauge
calculations in Section VII, we suggest a method in which the medium
internal energy can be approximately determined and subtracted by
using the information on the quark-gluon plasma of state.
Consequently, the $Q$-$\bar Q$ potential turns out to be a linear
combination of $U_1(R,T) $ and $F_1(R,T)$, with coefficients depending
on the equation of state.  In Section VIII, we use different
potentials to calculate the spontaneous dissociation temperatures for
various quarkonia and compare them with those obtained from lattice
gauge spectral function analyses in quenched QCD.  The dissociation
temperatures obtained with the proposed linear combination of $U_1$
and $F_1$ give the best agreement with those from the spectral
function analyses.  In Section IX, we show how the thermodynamical
quantities in full QCD with two flavors are parametrized.  The
dissociation temperatures for various heavy quarkonia for full QCD
with two flavors are obtained in Section X.  We introduce the quark
drip lines in quark-gluon plasma in Section XI.  We conclude and
summarize our discussions in Section XII.

\section{Thermodynamical quantities in lattice gauge calculations and 
Debye screening}

Thermodynamical quantities for a heavy quark pair in the color-singlet
state was studied by Kaczmarek and Zantow in quenched QCD and in full
QCD with two flavors \cite{Kac03,Kac05}.  They calculated $\langle tr
L({\bf r}/2) L^\dagger(-{\bf R}/2)\rangle$ and obtained the
color-singlet free energy $F_1({\bf R},T)$ from
\begin{eqnarray}
\label{free1}
\langle tr L({\bf R}/2) L^\dagger(-{\bf R}/2)\rangle = e^{-F_1({\bf
R},T)/kT}.
\end{eqnarray}
Here $tr L( {\bf R}/2) L^\dagger(-{\bf R}/2)$ is the trace of the
product of two Polyakov lines at ${\bf R}/2$ and $-{\bf R}/2$. The
free energy $F_1({\bf R},T)$, in the presence of the $Q$-$\bar Q$
pair, is measured relative to the free energy without the $Q$-$\bar Q$
pair.  The quark and the antiquark lines do not, in general, form a
close loop.  As a gauge transformation introduces phase factors at the
beginning and the end of an open Polyakov line, $\langle tr L( {\bf
R}/2) L^\dagger(-{\bf R}/2)\rangle$ is not gauge invariant under a
gauge transformation.  Calculations have been carried out in the
Coulomb gauge which is the proper gauge to study bound states.

From the free energy $F_1$, Kaczmarek and Zantow \cite{Kac03,Kac05}
calculated the internal energy $U_1$ using the statistical identity
\begin{eqnarray}
\label{FF}
U_{1} (R,T)=F_1(R,T)+TS_1(R,T),
\end{eqnarray}
where $S_1(R,T)=-\partial F_1(R,T)/\partial T$ is the entropy of the
system in the presence of a color-singlet $Q$-$\bar Q$ pair and is
measured relative to the entropy of the system in the absence of the
$Q$-$\bar Q$ pair.

In order to extract the $Q$-$\bar Q$ potential from thermodynamical
quantities calculated in the lattice gauge theory, we need to
understand the behavior of the free energy $F_1(R,T)$, the internal
energy $U_1(R,T)$, and $T$ times the entropy, $TS_1(R,T)$, which we
shall also abbreviatingly call ``the entropy'' for simplicity of
nomenclature.  As these three quantities are related by Eq.\
(\ref{FF}) we need to find out the behavior of only two of these three
quantities.  Following the terminology used in lattice gauge
calculations \cite{Kac05}, $U_1$, $F_1$, and $TS_1$ are defined as
being measured relative to their corresponding quantities in the
absence of $Q$ and $\bar Q$.

\begin{figure} [h]
\includegraphics[angle=0,scale=0.50]{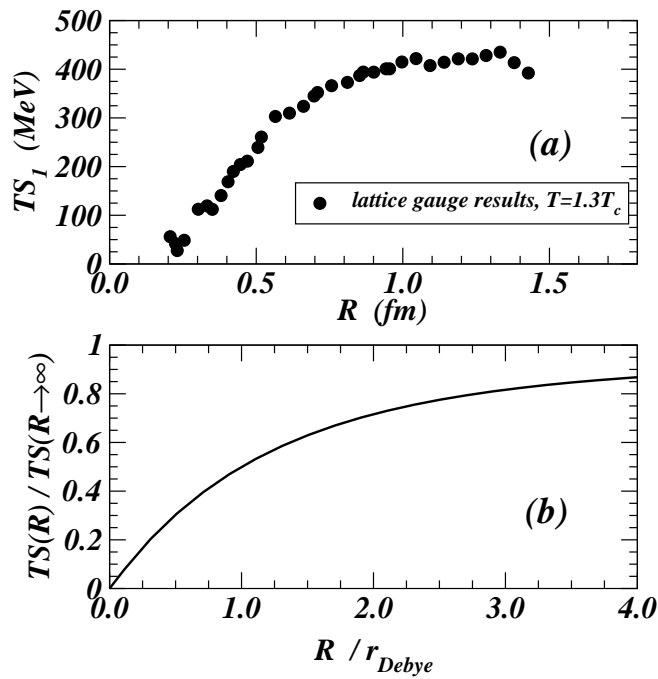}
\caption{($a$) The total entropy $TS_1(R,T)$ as a function of $R$ at
$T=1.3T_c$ from the lattice gauge calculations \cite{Kac05}.  ($b$)
The ratio $\eta(R)=TS(R)/TS(R\to \infty)$ (which is also equal to
$N(R)/N(R\to \infty)$ and $U(R)/U(R\to \infty)$) as a function of
$R/r_{\!_D}$ in a thermal medium under Debye screening.  }
\end{figure}

We can begin by studying the entropy $TS_1(R,T)$ of the system.  We
note that the lattice gauge calculations show that the total entropy
$TS_1(R)$ increases as a function of $R$, and saturates after the
separation reaches a large value of $R$ \cite{Kac05} as shown in
Fig. 1($a$).  What does such a behavior tell us about the response of
the medium particles to the presence of the external color sources of
$Q$ and $\bar Q$ in a thermal bath?

In the system under consideration, the system consists of the $Q$, the
antiquark $\bar Q$ and the quark-gluon plasma.  For simplicity we can
examine the quenched case for which the quark-gluon plasma is assumed
to consist of gluons only.  The entropy of the system therefore comes
from the sum of the entropies of $Q$, $\bar Q$ and the gluons.
However, in the lattice gauge calculations in a thermal bath, the $Q$
and the $\bar Q$ are held fixed and do not contribute to the entropy
of the system.  The entropy of the system $TS_1$ comes entirely from
the gluons.  By fixing a temperature and focusing our attention at the
state of thermal equilibrium in lattice gauge calculations, the gluons
are in a grand canonical ensemble in contact with the thermal bath.  The
content and thermodynamical properties of the gluons are determined by
the condition of thermal equilibrium at the fixed temperature of the
thermal bath.  The observed behavior of the entropy $TS_1(R)$ as a
function of $R$ in Fig.\ 1$(a)$ suggests that the gluon entropy content
increases as the separation $R$ increases, and the entropy saturates
when $R$ reaches a certain limit.

As we know from the work of Landau and Belenkii \cite{Lan53}, the
entropy content of a system is closely correlated with the number
content of the particles in the system. The behavior of the entropy
suggests that the number of gluons increases as the separation $R$
increases.  How do we understand such a behavior?  If the number of
gluons increases as a function of $R$, what happens to the internal
energy content of the gluons as a function of $R$, and how does that
affect the total internal energy and the potential between $Q$ and
$\bar Q$ we wish to extract?

\section{ Analogous problem of Debye screening in QED}

The QCD problem of a quark and an antiquark in the presence of the
actions of the gluons is so complicated that it is worth having even a
good understanding of the puzzling behavior in an analogous, but not
identical, problem of a static positive charge $Q$ and another
negative charge $\bar Q$ under the action of charged medium particles
in QED.  We would like to ask whether there is a similar behavior of
the number and the entropy of medium particles as the separation $R$
between the positive charge $Q$ and the negative charge $\bar Q$
increases.  If the number of medium particles increases as a function
of $R$, what happens to the internal energy content of the system and
the medium particles as a function of $R$, and what is the
relationship between the potential between $Q$ and $\bar Q$ and total
internal energy we wish to extract?

Accordingly, we study the simple system of Debye screening and
consider a $Q$ with charge $+q$ held fixed at $-{\bf R}/2$ and a $\bar
Q$ with a charge $-q$ at ${\bf R}/2$, in a medium of massless fermions
with charges $e_{\pm}=\pm q$, in a thermal bath of temperature $T$ in
a grand canonical ensemble.  Any pair of charged particles with
charges $e_1$ and $e_2$ separated by a distance ${\bf r}$ are assumed
to interact with a Coulomb interaction $e_1e_2/|{\bf r}|$.  To make
the problem simple, we shall assume the attainment of local thermal
equilibrium.

When the external charges $Q$ and $\bar Q$ are introduced into the
medium, the medium fermions will re-arrange themselves in both
coordinates and momenta to reach a new local thermal equilibrium.  The
total number of the medium particles in the system is not a constant
of motion but is determined by the condition of thermal equilibrium,
maintained by the thermal bath.

When the external charges $Q$ and $\bar Q$ are separated by a distance
${\bf R}$ at a temperature $T$, the self-consistent re-arrangement of
the medium charged fermion particles leads to a potential $V({\bf
r},{\bf R})$ at a point ${\bf r}$ and under a local thermal
equilibrium, the momentum distribution of the medium particles at
${\bf r}$ in the Born-Oppenheimer treatment is given by
\begin{eqnarray}
\label{fd}
f_{\pm}({\bf r},{\bf p},{\bf R})=\frac{1}
{\exp\{ [p+e_{\pm}V({\bf r},{\bf R})]/T\}+1}.
\end{eqnarray}
Here and henceforth, the
$+$ and $-$ subscripts designate quantities for the positive and
negative medium particles respectively.

From the above Wigner function distribution, we can obtain various
thermodynamic quantities.  The integration of the Wigner function over
all momenta gives the spatial number density distribution
$n_{\pm}({\bf r},{\bf R})$ at ${\bf r}$, when $Q$ and $\bar Q$ are
separated by ${\bf R}$,
\begin{eqnarray}
n_{\pm}({\bf r},{\bf R})=\frac {g}{2\pi^2} \int p^2 ~dp ~f_{\pm}({\bf
r},{\bf p},{\bf R}),
\end{eqnarray}
where $g$ is the degeneracy of the levels.  We can consider the high
temperature case for which it is useful to expand various quantities
as a power series of $e_{\pm} V(({\bf r},{\bf R})/T$.  Up to the
second order in $[e_{\pm}V({\bf r},{\bf R})/T]^2$, the medium particle
density is
\begin{eqnarray}
\label{eqn}
n_{\pm}({\bf r},{\bf R})=n_{\pm}^0 \left \{
1 - a_1 [e_{\pm} V({\bf r},{\bf R})/T] + a_2 [e_{\pm} V({\bf r},{\bf R})/T]^2 \right \},
\end{eqnarray}
where
\begin{eqnarray}
n_{\pm}^0= \frac{gT^3}{2\pi^2} \frac {3}{4} \zeta(3) \Gamma (3),
\end{eqnarray}
\begin{eqnarray}
\label{a1}
a_1= \frac{\frac{1}{2}\zeta(2)} {\frac{3}{4}\zeta(3)}=0.91233,
\end{eqnarray}
and
\begin{eqnarray}
\label{a2}
a_2= \frac{\frac{1}{2}\zeta(1)} {\frac{3}{4}\zeta(3)}=0.3845.
\end{eqnarray}
In passing, we note that if the medium particles obeys Boltzmann
statistics, the coefficients $a_1$ and $a_2$ would equal $a_1=1$ and
$a_2=0.5$, as it follows from the expansion of the well-known
Boltzmann distribution $n_{\pm}({\bf r},{\bf R})=n_{\pm}^0
\exp\{-e_{\pm} V({\bf r},{\bf R})/T \}$ for Boltzmann particles in an
external field.  The values of the $a_1$ and $a_2$ coefficients for
Boltzmann statistics differ only slightly from the corresponding
values in (\ref{a1}) and (\ref{a2}) for fermion particles.

For the Fermi-Dirac medium particles, the entropy density at ${\bf r}$, 
when $Q$ and $\bar Q$ are separated by ${\bf R}$, is given by 
\begin{eqnarray}
\sigma_{\pm}({\bf r},{\bf R})=\frac {g}{2\pi^2} \int p^2 ~dp 
\left \{ - f_{\pm} ~\ln f_{\pm}  - (1- f_{\pm} )~\ln (1- f_{\pm} ) \right \}.
\end{eqnarray}
Upon substituting the Fermi-Dirac distribution of Eq.\ (\ref{fd}) into
the above equation, we find
\begin{eqnarray}
\sigma_{\pm}({\bf r},{\bf R}) &=&\frac {g}{2\pi^2} \int p^2
~dp f_{\pm} ({\bf r},{\bf p},{\bf R})
 ~\left \{ \frac {4p}{3} + e_{\pm}V({\bf r},{\bf R}) \right
\}~\biggl / ~T \nonumber \\ 
\label{eqs0}
& \equiv &\langle \frac {4p}{3} + e_{\pm}V({\bf r},{\bf R})
\rangle~/~T,
\end{eqnarray}
and we obtain
\begin{eqnarray}
\label{eqs}
\sigma_{\pm}({\bf r},{\bf R})=\sigma_{\pm}^0 
\left \{
1 - b_1 [e_{\pm} V(({\bf r},{\bf R})/T] + b_2 [e_{\pm} V({\bf r},{\bf R})/T]^2 \right \},
\end{eqnarray}
where
\begin{eqnarray}
\sigma_{\pm}^0= \frac{gT^3}{2\pi^2} \frac {7}{8} \zeta(4) \frac{4}{3}\Gamma (4),
\end{eqnarray}
\begin{eqnarray}
b_1= \frac{\frac{3}{4}\zeta(3)[\frac {4}{3}\Gamma(4)-\Gamma(3)]} 
{\frac{7}{8}\zeta(4)\frac {4}{3}\Gamma(4) }=0.7139,
\end{eqnarray}
and
\begin{eqnarray}
b_2= \frac{\frac{1}{2}\zeta(2)[\frac {4}{3}\frac{\Gamma(4)}{2} -\Gamma(3)]} 
{\frac{7}{8}\zeta(4)\frac {4}{3}\Gamma(4) }=0.2171.
\end{eqnarray}
Here we have purposely written the numerators of the $b_1$ and $b_2$
coefficients as a difference where the first term comes from $\langle
4p/3\rangle$ and the second term comes from $\langle e_\pm V \rangle$ of
Eq. (\ref{eqs0}).

The evaluation of various thermodynamical quantities requires the
knowledge of $V({\bf r},{\bf R})$.  To determine $V({\bf r},{\bf R})$
self-consistently, we have the charge density at the point ${\bf r}$,
when $Q$ and $\bar Q$ are separated by ${\bf R}$,
\begin{eqnarray}
\rho_{\rm total}({\bf r},{\bf R}) = q \delta({\bf r}+\frac {{\bf
R}}{2}) -q \delta({\bf r}-\frac {{\bf R}}{2}) +e_+ n_+ ({\bf r},{\bf
R}) + e_- n_- ({\bf r},{\bf R}).
\end{eqnarray}
Using the number density distribution given in Eq.\ (\ref{eqn}), the
charge density, up to the second power in $e_\pm V({\bf r},{\bf
R})/T$, becomes
\begin{eqnarray}
\rho_{\rm total}({\bf r},{\bf R})
= q  \delta({\bf r}+\frac {{\bf R}}{2}) -q \delta({\bf r}-\frac {{\bf R}}{2})
       - n_0 a_1 q^2 V({\bf r},{\bf R}), 
\end{eqnarray}
where $n_0=n_+^0+n_-^0$ and the zeroth-order and second-order terms of
 Eq.\ (\ref{eqn}) cancel out on account of the presumed charge
 neutrality of the system for which $n_+^0=n_-^0$.  The Poisson
 equation for the potential is then given by
\begin{eqnarray}
\nabla_{\bf r}^2 V ({\bf r},{\bf R}) = -4\pi \left \{q ~ \delta({\bf r}+\frac
{{\bf R}}{2}) -q ~\delta({\bf r}-\frac {{\bf R}}{2}) - n_0 a_1 q^2
V({\bf r},{\bf R})\right \},
\end{eqnarray}
which has the solution

\begin{subequations}
\label{sol}
\begin{eqnarray}
V ({\bf r},{\bf R})
&=& \frac {q e^{-\mu r_+}}{r_+} - \frac {q e^{-\mu r_-}}{r_-},
\label{solv}\\
r_{\pm} &=& |{\bf r}\pm {\bf R}/2|, 
\\
\mu &=& \sqrt{ \frac {4 \pi n_0 a_1 q^2 }{T}}=\frac {1}{r_{\!_D}},
\end{eqnarray}
\end{subequations}
where $\mu$ is the Debye mass and $r_{\!_D}$ is the Debye screening length.

\section{Variation of number density and entropy density with $R$ in
  Debye screening}

The simple solution $V ({\bf r},{\bf R})$ in Eq.\ (\ref{sol}) of the
last section allows us to have a profile of the self-consistent medium
particle number density and entropy density in all spatial points at
local thermal equilibrium at $T$.  In Eqs.\ (\ref{eqn}) and
(\ref{eqs}), the coefficients of $a_1$, $a_2$, $b_1$ and $b_2$ are all
positive.  For positive medium particles, the first-order increment in
the number density $n_+({\bf r},{\bf R})$ and $\sigma_+({\bf r},{\bf
R})$, [also $u_+({\bf r},{\bf R})$ in Eq.\ (\ref{equ31})] are therefore
measured by $[-e_+V({\bf r},{\bf R})/T]$ illustrated in Fig.\ 2 as a
function of $\rho/r_{\!_D}$ and $z/r_{\!_D}$.  As one observes, the
first-order contributions, given by $-e_+V({\bf r},{\bf R})/T$,
represent a depletion for the positive medium particles near the
positive static charge $Q$ at $-{\bf R}/2$, and an enhancement near
the negative static charge $\bar Q$ at ${\bf R}/2$.  The degree of
depletion and the degree of enhancement are equal and opposite to each
other.  When we sum over all spatial points, the sum of the
first-order depletion and enhancement cancel each other to give a zero
total contribution.

\begin{figure} [h]
\includegraphics[angle=0,scale=0.75]{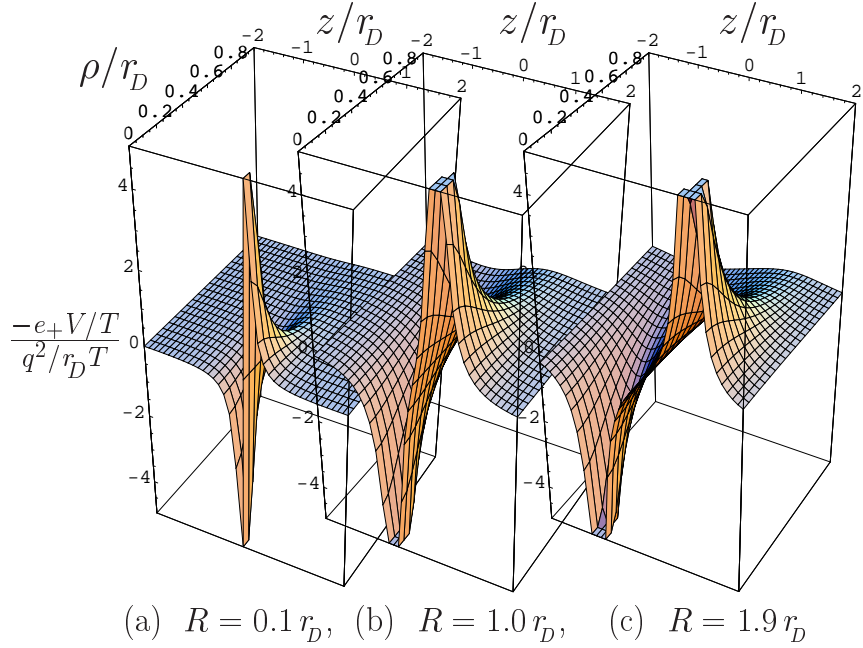}
\caption{ The first-order term $[-e_+V(\rho z,T)/T]$ in units of
$q^2/r_{\!_D}$ in Eqs.\ (\ref{eqn}), (\ref{eqs}), and (\ref{equ31})
that contributes to the increments in $n_+(\rho z,R), \sigma_+(\rho
z,R),$ and $ u_+(\rho z,R)$, as a function of the spatial coordinates
$\rho/r_{\!_D}$ and $z/r_{\!_D}$.  Fig. 2($a$) is for $R=0.1
r_{\!_D}$, Fig. 2($b$) for $R= 1.0r_{\!_D}$, and Fig. 2($c$) for $R=1.9
r_{\!_D}$.  }
\end{figure}

The second-order contributions are proportional to $[e_+V({\bf r},{\bf
R})/T]^2$ and are always positive.  They are illustrated in Fig.\ 3 as
a function of $\rho/r_{\!_D}$ and $z/r_{\!_D}$ for different $Q$-$\bar
Q$ separations.  They always enhance the number density and the entropy
density.  The enhancement is small when the two static charges are
close together in Fig. 3($a$), as there is a substantial cancellation
of the two terms in Eq.\ (\ref{sol}$a$).  The enhancement reaches a
constant value when the static charges $Q$ and $\bar Q$ reaches a
separation of 1-2 units of the Debye screening length as shown in
Fig. 3$b$ and 3$c$.
 
If one integrates over all spatial points to obtain the total number
of positive charge medium particles, one finds that the integration
over the first-order term, $\int d{\bf r} [-e_+V({\bf r},{\bf R})/T]$,
is zero because the depletion cancels the enhancement.  However, the
second-order contributions always give a positive contribution, and
the number of positive medium particles, measured relative to its
corresponding quantity in the absence of $Q$ and $\bar Q$, is given by
\begin{eqnarray}
\label{nr}
 N_+(R) = \int d{\bf r} \left \{ n_+({\bf r},{\bf R}) -
n_+^0\right \} = n_+^0 \int d{\bf r} ~a_2 [e_+V({\bf r},{\bf R})/T]^2.
\end{eqnarray}
We can write the above as
\begin{eqnarray}
\label{ninf}
\frac{ N_+(R)} { N_+(R\to \infty)} =\eta(R)
\end{eqnarray}
where
\begin{eqnarray}
\label{eta}
\eta(R)=\frac {1}{4\pi}\int 
 ~d{\bzeta}~ \left [ \frac {e^{-\zeta_+}}{\zeta_+} 
                   - \frac {e^{-\zeta_-}}{\zeta_-}
\right ]^2 ,
\end{eqnarray}
with ${\bzeta}={\bf r}/r_{\!_D}$, $\zeta_\pm=|{\bf r}\pm {\bf
R/2}|/r_{\!_D}$, $\eta(0)=0$, and $\eta(R\to \infty)=1$.  In Eq.\
(\ref{ninf}), $ N_+(R\to \infty)$ is the increment in the number of
positive medium particles when $Q$ and $\bar Q$ are far separated,
\begin{eqnarray}
 N_+(R\to \infty) = n_+^03 a_2 \left ( \frac {q^2}{r_{\!_D} T} \right ) ^2 
\frac {4\pi }{3} r_{\!_D}^3.
\end{eqnarray}
The increase in the total number of positive medium particles is small
when $Q$ and $\bar Q$ are close together, and the increase saturates
when $R$ reaches a few units of the Debye screening length
$r_{_{\!_D}}$.  For our charge-neutral system, we obtained from
Eq.\ (\ref{eqn}) in a similar way $N_-(R)=N_+(R)$.

\begin{figure} [h]
\centering
\includegraphics[angle=0,scale=0.75]{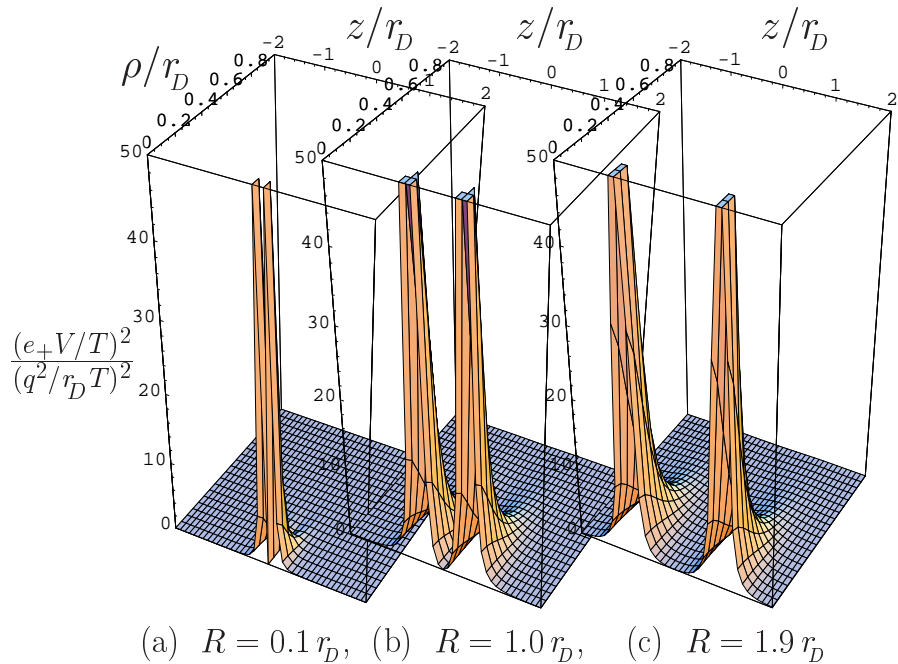}
\caption{ The second-order term $(e_+V(\rho z,R)/T)^2$ in units of
$(q^2/r_{\!_D}T)^2$ in Eqs.\ (\ref{eqn}), (\ref{eqs}), and (\ref{equ31})
that contributes to $n_+(\rho z,R)$, $\sigma_+(\rho z,R),$ and $
u_+(\rho z,R)$, as a function of the spatial coordinates
$\rho/r_{\!_D}$ and $z/r_{\!_D}$. Fig. 3($a$) is for $R=0.1 r_{\!_D}$,
Fig. 3($b$) for $R= 1.0r_{\!_D}$, and Fig. 3($c$) for $R=1.9 r_{\!_D}$.
}
\end{figure}

Similarly, the entropy density of the medium particles is depleted
near the static charge of the same sign, and is enhanced in the
vicinity of the static charge of the opposite sign.  When integrated
over all spatial points, the depletion and the enhancement cancel to
the first order.  The second-order contributions always give a positive
total entropy,
\begin{eqnarray}
T S_\pm(R) = T\int d{\bf r} \left \{ \sigma_\pm({\bf r},{\bf R}) -
\sigma_\pm^0\right \} = T\sigma_\pm^0 \int d{\bf r} ~b_2 
[e_\pm V({\bf r},{\bf R})/T]^2,
\end{eqnarray}
which is small at small $R$ and saturates at large $R$.  We can write
the above as
\begin{eqnarray}
\frac{T S_\pm(R)} {T S_\pm(R\to \infty)} =\eta(R)
\end{eqnarray}
where
\begin{eqnarray}
T S_\pm(R\to \infty) = T\sigma_\pm^03 b_2 \left ( \frac {q^2}{r_{\!_D} T} \right ) ^2 
\frac {4\pi }{3} r_{\!_D}^3.
\end{eqnarray}
A comparison of $ N_\pm$ and $ TS_\pm$ shows that 
\begin{eqnarray}
\frac{ N_\pm(R)} { N_\pm(R\to \infty)} =
\frac{T S_\pm(R)} {T S_\pm(R\to \infty)}=\eta(R) .
\end{eqnarray}
If we define $N(R)=N_+ + N_-(R)$, and $TS(R)=TS_+(R)+TS_-(R)$, the
increment of the total number and entropy of the medium particles due
to the presence of $Q$ and $\bar Q$ (measured relative to the
corresponding quantities in the absence of $Q$ and $\bar Q$), is
\begin{eqnarray}
\label{eqeq}
\frac{ N(R)} { N(R\to \infty)} =
\frac{T S(R)} {T S(R\to \infty)}=\eta(R) .
\end{eqnarray}
Thus, the ratios ${ TS(R)}/ { TS(R\to \infty)}$, ${N_\pm(R)}/ {
N_\pm(R\to \infty)}$, and ${ TS (R)}/ { TS (R\to \infty)}$ [also
${U(R)}/ { U(R\to \infty)}$ as we shall see in the next section] behave in
the same way as a function of $R$.  We show the behavior of $\eta(R)={
TS(R)} /{ TS(R\to \infty)}$ for the Debye screening case in Fig\
1($b$) and it has the same shape as $TS_1(R)$ obtained in the lattice
gauge calculations shown in Fig.\ 1($a$).

One can therefore understand that as the result of constraining the
system to be in contact with a thermal bath in the grand canonical
ensemble, the medium particle numbers and entropy increase with
increasing $R$ to maintain a thermal equilibrium until they saturate
at large separation.  The thermal bath is therefore a participant in
altering the content of the medium particles, when the $Q$ is
separating from the antiquark $\bar Q$. (See Section VI for a
discussion on the role of the thermal bath).

\section{Variation of the internal energy with $R$ in Debye Screening}

Previously in quenched QCD, we find a relationship between the total
internal energy and the $Q$-$\bar Q$ potential \cite{Won05}.  As the
problem of non-perturbative QCD so intrinsically complicated, a
thorough understanding of such a relationship in an analogous, but not
identical, problem in QED is worth having.  We are therefore motivated
to examine the mechanism of Debye screening due to the medium
particles in QED.  Of particular interest is to see whether the
relationship between the total internal energy and the $Q$-$\bar Q$
potential in Debye screening in QED resembles a similar relationship
in lattice gauge theory obtained previously in QCD \cite{Won05}.

When $Q$ and $\bar Q$ separated by a distance $R$ are screened by
medium particles, the total internal energy $U_{\rm total}(R)$ of the
system of medium particles, $Q$, and $\bar Q$ is the sum of the
kinetic energy of the medium particles and the interaction energies of
the the medium particles, $Q$, and $\bar Q$, when the medium particles
have re-arranged themselves self-consistently,
\begin{eqnarray}
\label{Utotal}
U_{\rm total}(R)&=& \int d{\bf r} \left \{
\frac{g}{2\pi^2} \int p^2 dp
[f_+({\bf r},{\bf p},{\bf R}) + f_-({\bf r},{\bf p},{\bf R}) ] p
\right \}
\nonumber\\ &+& \frac{1}{2} \int d{\bf r} 
\left \{ e_+ \delta({\bf r}+\frac {{\bf R}}{2}) + e_- \delta({\bf r}-\frac {{\bf R}}{2})
       +e_+ n_+({\bf r},{\bf R}) +e_- n_-({\bf r},{\bf R}) \right \} 
V({\bf r},{\bf R})\\
&-&\int d{\bf r}
(u_+^0+u_-^0),\nonumber
\end{eqnarray}
where
$u_\pm^0$ is the internal energy density of the medium 
in the absence of $Q$ and $\bar Q$ given by
\begin{eqnarray}
u_{\pm}^0= \frac{gT^4}{2\pi^2} \frac {7}{8} \zeta(4) \Gamma (4),
\end{eqnarray}
and $U_{\rm total}(R)$ is measured relative to the total internal
energy of the system in the absence of $Q$ and $\bar Q$, $U_{\rm
medium}^0=\int d {\bf r}(u_+^0 + u_-^0)$.  Using the solution of
$V({\bf r},{\bf R})$ in Eq.\ (\ref{sol}) and excluding the infinite
energies of a point source acting on itself, Eq.\ (\ref{Utotal}) gives
\begin{eqnarray}
\label{Utotal1}
U_{\rm total}(R)=
-\frac {q^2 e^{-\mu R}}{R} - \frac{q^2}{r_{\!_D}}+\int d{\bf r} 
\left [ \{ u_+({\bf r},{\bf R})-u_+^0\} + \{ u_-({\bf r},{\bf R}) -u_-^0 \}
\right ], 
\end{eqnarray}
where $u_\pm({\bf r},{\bf R})$ are the internal energy density of the
positive and negative charged medium particles in the presence of $Q$
and $\bar Q$ given by
\begin{eqnarray}
\label{equ}
u_{\pm}({\bf r},{\bf R}) &=&\frac {g}{2\pi^2} \int p^2
~dp f_{\pm} ({\bf r},{\bf p},{\bf R})
\left \{ p + \frac{e_{\pm}V({\bf r},{\bf R})}{2} \right
\} \nonumber \\ 
&\equiv &\langle p + \frac{e_{\pm}V({\bf r},{\bf R})}{2}
\rangle  .
\end{eqnarray}
Using the local Fermi-Dirac distribution of Eq.\ ({\ref{fd}),
we obtain
\begin{eqnarray}
\label{equ31}
u_{\pm}({\bf r},{\bf R})=u_{\pm}^0 
\left \{
1 - c_1 [e_{\pm} V(({\bf r},{\bf R})/T] + c_2 [e_{\pm} V({\bf r},{\bf R})/T]^2 \right \},
\end{eqnarray}
where $c_1$ and $c_2$ are positive constants, 
\begin{eqnarray}
c_1= \frac{\frac{3}{4}\zeta(3)[\Gamma(4)-\frac{\Gamma(3)}{2}]} 
{\frac{7}{8}\zeta(4)\Gamma(4) }=0.7933,
\end{eqnarray}
and
\begin{eqnarray}
c_2= \frac{\frac{1}{2}\zeta(2)[\frac{\Gamma(4)}{2} 
-\frac{\Gamma(3)}{2}]} {\frac{7}{8}\zeta(4)}=0.2895.
\end{eqnarray}
A comparison of $u_\pm({\bf r},{\bf R})$ in Eq.\ (\ref{equ31}) with
$n_\pm({\bf r},{\bf R})$ and $u_{\pm}({\bf r},{\bf R})$ in Eqs.\
(\ref{eqn}) and (\ref{eqs}) indicates that the first-order and
second-order contributions to $u_\pm({\bf r},{\bf R})$ behave in the
same way as those of $n_\pm({\bf r},{\bf R})$ and $\sigma_\pm({\bf
r},{\bf R})$.  The first-order increment of the internal energy
density of the medium particles is suppressed near the static charge
of the same sign, and is enhanced near the static charge of the
opposite sign as shown in Fig. 2, while the second-order contributions
to the internal energy are always positive as shown in Fig. 3.  The
first-order contributions cancel each other, when they are integrated
over all the spatial points.  However, the second-order contributions
are always positive, and the integration of the second-order
contributions always yield a positive quantity.  The total medium
internal energy of the positive and negative medium particles,
measured relative to the corresponding quantities in the absence of
$Q$ and $\bar Q$, are
\begin{eqnarray}
\label{ur}
 U_\pm(R) = \int d{\bf r} \left \{ u_\pm({\bf r},{\bf R}) -
u_\pm^0\right \} = n_\pm ^0 \int d{\bf r} ~c_2 [e_\pm V({\bf r},{\bf R})/T]^2.
\end{eqnarray}
Relative to the total internal energy in the absence of $Q$ and $\bar
Q$, the total internal energy in the presence of $Q$ and $\bar Q$, is
\begin{eqnarray}
\label{equu}
U_{\rm medium} (R)-U_{\rm medium}^0
=U_+ (R)+U_-(R). 
\end{eqnarray}
Eq.\ (\ref{Utotal1}) can therefore be written as 
\begin{eqnarray}
\label{Utotalln}
U_{\rm total}(R)=
-\frac {q^2 e^{-\mu R}}{R} - \frac{q^2}{r_{\!_D}}
+U_{\rm medium} (R)-U_{\rm medium}^0
\end{eqnarray}
From Eqs.\ (\ref{ur}), (\ref{nr}), and (\ref{eta}),
the quantity $U_\pm (R)$ can be written as
\begin{eqnarray}
\frac{ U_\pm (R)} { U_\pm (R\to \infty)} =\eta(R),
\end{eqnarray}
where
\begin{eqnarray}
 U_\pm (R\to \infty) = u_\pm ^0 3 c_2 \left ( \frac {q^2}{r_{\!_D} T} \right ) ^2 
\frac {4\pi }{3} r_{\!_D}^3.
\end{eqnarray}
A comparison of $ N_\pm$ and $ TS_\pm$ shows that 
\begin{eqnarray}
\frac{ U_\pm(R)} { U_\pm(R\to \infty)} =
\frac{ N_\pm(R)} { N_\pm(R\to \infty)} =
\frac{ TS_\pm(R)} { TS_\pm(R\to \infty)}=\eta(R) .
\end{eqnarray}
Consequently, if we define
$U(R)=U_+(R)+U_-(R)$, we also have
\begin{eqnarray}
\label{ratio}
\frac{ U(R)} { U(R\to \infty)} =
\frac{ N(R)} { N(R\to \infty)} =
\frac{ TS(R)} { TS(R\to \infty)}=\eta(R) .
\end{eqnarray}
In this simple model of Debye screening, the ratios of $TS(R)/TS(R\to
\infty)$, $N(R)/N(R\to \infty)$ and $U(R)/U(R\to \infty)$ are equal
and their behavior is shown in Fig.\ 1($b$).  The entropy, total
number, and internal energy of the medium particles (relative to the
corresponding quantities in the absence of $Q$ and $\bar Q$) is zero
at $R=0$ and increases as a function of $R$ until they saturate when
the separation $R$ reaches a few units of the Debye screening length.

\section{Schr\" odinger equation for $Q$-$\bar Q$ system in Debye
  screening}

As the dynamics of a quark and an antiquark in a quark-gluon plasma is
very complicated, it is worth having a good understanding of the
mechanism of screening and its effects on the interaction between a
$Q$ and an antiquark $\bar Q$ in an analogous, but not identical,
problem.  We are therefore motivated to examine the mechanism of Debye
screening due to the medium particles in QED.  Of particular interest
is to obtain the relationship between the $Q$-$\bar Q$ potential in
the Hamiltonian and the total internal energy $U_{\rm total}$, in
order to find out whether this relationship resembles a similar
relationship in lattice gauge theory obtained previously in Ref.\
\cite{Won05}.

It is however somewhat tricky to determine the Hamiltonian for the $Q$
and $\bar Q$ system in a medium under Debye screening (and
analogously, but not identically, under color-charge screening in
QCD).  We can follow the basic principles of statistical physics as
described by Landau and Lifshitz \cite{Lan59}.  Accordingly, we start
with a {\it closed} system of a heavy $Q$ and $\bar Q$ with medium
particles and consider a small ``subsystem'' $S$ that contains the
$Q$, the $\bar Q$, and the medium.  As the number of medium particles
of the whole closed system is very large, the number of medium
particles contained in the small subsystem $S$ can still be very
large, and a statistical description of this small subsystem $S$ is
applicable.  This subsystem $S$ is not a closed system and it
undergoes all kinds of interaction and medium particle exchanges with
the complementary part $S'$ of the whole system.  We can describe this
subsystem $S$ to be in contact with a very large complementary part
$S'$, which we can call a ``thermal bath'' in this connection.  The
QED system discussed in the last few sections or the QCD system in
lattice gauge calculations, (the so-called ``system'' of $Q$, $\bar
Q$, and medium particles in contact with a thermal bath), corresponds
in actual fact to the ``subsystem'' $S$ out of the whole closed system
$S+S'$.

Under a perturbation of the subsystem $S$ away from thermal
equilibrium such as occurs in the displacement of $Q$ relative to
$\bar Q$, the medium in the subsystem $S$ will respond to the
perturbation and will relax to a new state of thermal equilibrium
after a certain relaxation time, $t_{\rm relax}(S)$. For example, from
the results concerning the medium entropy and number contents in the
subsystem $S$ as a function of the separation between $Q$ and $\bar Q$
obtained in the last few sections (Fig. 1), we know that under a
displacement of the relative separation of the $Q$ and $\bar Q$, the
medium particles will exchange between $S$ and the thermal bath $S'$
in order to make the subsystem $S$ under thermal equilibrium.  In this
case, the relaxation time $t_{\rm relax}(S)$ corresponds to the
exchange of medium particles through imaginary boundaries between $S$
and the thermal bath $S'$.  Relaxation time grows smaller as the
subsystem $S$ decreases in size \cite{Lan59}.  For a small subsystem
$S$, this relaxation time $t_{\rm relax}(S)$ can be very short.  On
the other hand, for heavy $Q$ and $\bar Q$ in the subsystem, the
period of $Q$-$\bar Q$ relative motion, $t_{Q\bar Q}$, can be
relatively long because of the large mass of $Q$ and $\bar Q$.  The
period $t_{Q\bar Q}$ can be so much greater than the medium relaxation
time $t_{\rm relax}(S)$, $t_{Q\bar Q}>>t_{\rm relax}(S)$, that the
medium can be approximately considered as reaching a state of thermal
equilibrium approximately instantaneously, at any time during the
(supposedly slow) motion of the heavy quark $Q$ and antiquark $\bar
Q$.  For the medium particles, this is just the Born-Oppenheimer
approximation for the description of the states of the medium
particles in the subsystem $S$, as presented in the last few sections
and used in lattice gauge calculations to obtain the medium particle
configurations in QCD.

In what sense can energy and entropy be considered conserved under a
periodic motion of the $Q$ and $\bar Q$ in the medium?  The whole
closed system consists of the subsystem $S'$ and $S$, and the
subsystem $S$ is not a closed system.  From the results of the last
few sections, we know that if we move the $Q$ closer relative to the
$\bar Q$ in the subsystem $S$, then the medium particles (and its
entropy and energy contents) will move from the subsystem $S$ into the
complementary part $S'$ so as to maintain thermal equilibrium in the
subsystem $S$.  When we move the $Q$ farther relative to the $\bar Q$,
then the medium particles, entropy, and energy contents will move from
the complementary system $S'$ back into the subsystem $S$ so as to
maintain thermal equilibrium in the subsystem $S$.  For slow periodic
motion of $Q$ and $\bar Q$, the motion can be so slow that the
exchange of medium particles between $S$ and $S'$ can be approximated
as taking place with no excitation of the medium.  In this sense, the
idealized periodic motion of $Q$ and $\bar Q$ can thus be {\it
adiabatic} in the lowest order with an ``adiabatic'' exchanging the
energy content and the entropy content of the medium particles back
and forth between the subsystem $S$ and the complementary system $S'$.
Additional interactions of the bound periodic $Q$-$\bar Q$ states with
the medium particles that lead to the non-adiabatic excitation of both
objects can then be considered in higher-order approximations.

Thus, in this adiabatic picture of $t_{Q\bar Q}>>t_{\rm relax}(S)$,
the $Q$ and $\bar Q$ experience the interactions from all medium
particles which adjusts themselves (within a relaxation time which is
taken to be so small as to be approximately instantaneous) at all
instances of the dynamical motion of the $Q$ and $\bar Q$.  The
potential energy of the $Q$ and $\bar Q$ at a separation $r$ in the
Hamiltonian for the $Q$ and $\bar Q$ is half of the integral of the
product of the local point charges of the $Q$ and $\bar Q$ with their
local potentials $V({\bf r},{\bf R})$ arising from a self-consistent
rearrangement of all particles when $Q$ and $\bar Q$ are at the
separation ${\bf R}$, excluding the infinite self-energy
contributions.  In addition to the potential energy of the $Q$ and
$\bar Q$, the Hamiltonian for the $Q$-$\bar Q$ system consists also of
the kinetic energy of $Q$ and $\bar Q$.  The Hamiltonian for the
$Q$-$\bar Q$ system is therefore given by
\begin{eqnarray}
\label{H}
H&=& \frac{{\bf p}_Q^2 }{2 m_Q} + \frac{{\bf p}_{\bar Q}^2 }{2 m_{\bar Q}}
+ \frac{1}{2} \int d{\bf r} 
\left [ e_+ \delta({\bf r}+\frac {{\bf R}}{2}) 
+ e_- \delta({\bf r}-\frac {{\bf R}}{2}) \right ]V({\bf r},{\bf R})
\nonumber\\
&=& \frac{{\bf p}_Q^2 }{2 m_Q} +\frac{{\bf p}_{\bar Q}^2 }{2 m_{\bar Q}}
+  \frac{1}{2}[ e_+ V({\bf r},{\bf R})|_{{\bf r}=-{\bf R}/{2}} 
+ e_- V({\bf r},{\bf R})|_{{\bf r}={\bf R}/{2}} ].
\end{eqnarray}
Upon making the change of the variables from
${\bf p}_Q$ and ${\bf p}_{\bar Q}$ to the center-of-mass momentum
${\bf P}_{\rm CM}={\bf p}_Q+{\bf p}_{\bar Q}$ and the relative
momentum ${\bf p}_{R}=({\bf p}_Q-{\bf p}_{\bar Q})/2$, and using the
solution $V({\bf r},{\bf R})$ of Eq.\ (\ref{sol}), we find from the
above equation
\begin{eqnarray}
\label{eq42}
H= \frac{{\bf P}_{\rm CM}^2 }{2 (m_Q+m_{\bar Q})} 
+\frac{{\bf p}_{R}^2 }{2\mu_{\rm red}}
 - \frac{q^2 e^{-\mu R}}{R} - \frac{q^2}{r_{\!_D}},
\end{eqnarray}
where $\mu_{\rm red}=m_Q m_{\bar Q}/(m_Q+m_{\bar Q})$ is the reduced
mass.  The Hamiltonian then separates into $H=H_{\rm CM}+H_R$ where
$H_{\rm CM}$ is the Hamiltonian for the free motion of the composite
two-body system and $H_R$ is the Hamiltonian for the relative motion
of $Q$ and $\bar Q$,
\begin{eqnarray}
\label{HR}
H_R = \frac{{\bf p}_{R}^2 }{2\mu_{\rm red}} - \frac{q^2 e^{-\mu R}}{R}
- \frac{q^2}{r_{\!_D}}
 \equiv  \frac{{\bf p}_{R}^2 }{2\mu_{\rm red}} +
U_{Q\bar Q}(R)
\end{eqnarray}
From the above equation, we recognize that the potential for the
$Q$-$\bar Q$ system under screening by the medium, $U_{Q\bar Q}(R)$,
is given by
\begin{eqnarray}
\label{uqqr}
U_{Q\bar Q}(R)= - \frac{q^2 e^{-\mu R}}{R} - \frac{q^2}{r_{\!_D}}.
 \end{eqnarray}
The ${Q\bar Q}$ potential $U_{Q\bar Q}(R)$ is just the Debye screening
potential plus an $R$-independent constant term.

Based on the ``adiabatic'' picture of the motion of $Q$ and $\bar Q$,
the Hamiltonian formulated in Eq.\ (\ref{H}) indeed gives correctly
the Hamiltonian with the Debye screening potential. The
$R$-independent term $- {q^2}/{r_{\!_D}}$ is also an important part of
the screening contribution.  We note that if we expand the $Q$-$\bar
Q$ Hamiltonian for the case of small $\mu$ representing the screening
effects, then Eq.\ (\ref{HR}) becomes,
\begin{eqnarray}
H_R \sim \frac{{\bf p}_{R}^2 }{2\mu_{\rm red}} - \frac{q^2 (1-\mu
  R)}{R} - \frac{q^2}{r_{\!_D}}= \frac{{\bf p}_{R}^2 }{2\mu_{\rm
    red}}- \frac{q^2}{R},
\end{eqnarray}
which is the same Hamiltonian as that of the unscreened case.  Thus,
we reach the interesting result that in the lowest order of the
screening parameter $\mu=1/r_{\!_D}$, a properly calibrated
Hamiltonian of a system under screening, with the shift of the level
of the potential, $-q^2/r_D$, is the same Hamiltonian without
screening.  In practical terms, if we calculate the mass of a
$Q$-$\bar Q$ system without screening, we expect that within the
lowest order of the screening parameter $\mu$, the mass eigenvalue of
the system to be nearly unchanged when screening is present.
Numerical calculations of the mass of bound $L=0$ charmonium using the
potential model of Ref.\ \cite{Won05} indeed shows that the absolute
value of the charmonium $L=0$ mass changes only very slightly as a
function of temperature up to 1.5$T_c$ \cite{Won06b}.

From the above discussions in the simple case of Debye screening, we
observe that the potential between $Q$ and $\bar Q$, $U_{Q\bar Q}(R)$,
differs from the total internal energy $U_{\rm total}(R)$.  Because of
Eqs.\ (\ref{uqqr}) and (\ref{Utotalln}), they are related by
\begin{eqnarray}
U_{Q\bar Q}(R)=U_{\rm total}(R)-[U_{\rm medium} (R)-U_{\rm medium}^0 ].
 \end{eqnarray}
It is therefore necessary to subtract out the change of the medium
internal energies [$U_{\rm medium}(R)-U_{\rm medium}^0$] from the
total internal energy $U_{\rm total}(R)$ to obtain the $Q$-$\bar Q$
potential $U_{Q\bar Q}(R)$ in the grand canonical ensemble.  This
conclusion for Debye screening supports a similar conclusions in the
analogous lattice gauge theory, where we have proved in Eq.\ (11) of
Ref.\ \cite{Won05},
\begin{eqnarray}
\label{equqq}
U_{Q\bar Q}^{(1)}(R,T)=U_1(R,T)-[U_g^{(1)} (R,T)-U_{g0}(T)].
\end{eqnarray} In the above equation, the superscript $(1)$ refers to
the color-singlet state of $Q$ and $\bar Q$, and $[U_g^{(1)}
(R,T)-U_{g0}(T)]$ is the increment of gluon energy due to the presence
of $Q$ and $\bar Q$.

From the above analysis, we conclude that the relationship of Eq.\
(\ref{equqq}) in Ref.\ \cite{Won05} is a rather general result for
heavy particles under screening in the grand canonical ensemble.

     We can understand the results of Eq. (48) [or similarly (47)]
     from another viewpoint.  In a standard description of a $(Q\bar
     Q)$ in a medium, we simplify the dynamics by considering first
     the $(Q \bar Q)$ states and the deconfined medium states
     separately as independent unperturbed states.  We then include
     their mutual excitations as perturbative couplings. Thus, in the
     lowest-order description without perturbative couplings, the
     $(Q\bar Q)$ states should be obtained without the excitation of
     the medium states of deconfined real gluons and vice versa. In
     the quenched approximation, the deconfined real gluons in the
     quark-gluon plasma in the Feynman diagram language are those
     represented by lines with external legs.  The change of the
     medium internal energies, $[U_{g}^{(1)} (R,T)-U_{g0}(T)]$ in
     Eq. (48), represents the excitation of the internal energy states
     of the deconfined real gluon medium when the separation between
     the $Q$ and the $\bar Q$ changes in the grand canonical ensemble.
     As the unperturbed states of the $Q$-$\bar Q$ relative motion
     should be calculated without the excitation of the medium states
     of deconfined real gluons, we therefore need to subtract the
     change of the real gluon internal energy from the total internal
     energy in Eq. (48) to obtain the $Q$-$\bar Q$ potential $U_{Q\bar
     Q}^{(1)}(R,T)$, when the separation between the $Q$ and the $\bar
     Q$ changes.

     In the quenched approximation, the subtraction of this change in
     the internal energy of real gluons as a function of $R$ does not
     mean that both the real and the virtual gluon degrees of freedom
     are frozen.  Only the real gluon excitation energy degrees of
     freedom are frozen when we calculate the unperturbed $Q$-$\bar Q$
     bound states of relative motion, for reasons we have just given.
     On the other hand, virtual gluons, which in the Feynman diagram
     language are represented by gluon lines with the two end points
     of each line joining onto other quarks and gluons, change their
     configurations as the separation $R$ between the $Q$ and the
     $\bar Q$ changes.  These virtual gluons mediate the interaction
     between the $Q$ and the $\bar Q$.  The changes in the virtual
     gluon configurations modify the interaction between the $Q$ and
     the $\bar Q$, resulting in the screening of the $Q$-$\bar Q$
     potential.  This type of virtual gluon excitation is not frozen
     and is included in the calculation of the $Q$-$\bar Q$ potential
     and the evaluation of bound states of the $Q$-$\bar Q$ relative
     motion.

     The reconfiguration of these virtual gluons can take place in
     either an adiabatic manner or in the opposite ``diabatic" manner
     depending on whether the time scale of the relaxation of these
     virtual gluons is short compared to the time scale of the period
     of the $Q$-$\bar Q$ relative motion.  As we explained earlier,
     the adiabatic description is appropriate for very heavy quark
     pairs when the relative motion of the $Q$ and the $\bar Q$ is
     slow.  This is indeed confirmed by the successful identification
     of the lattice free energy for a static $Q$-$\bar Q$ pair as the
     heavy quark $Q$-$\bar Q$ potential, $U_{Q\bar Q}^{(1)}(R)$ at
     $T=0$ [Bali et al.  Phys. Rev. D56 2566 (1997)].  For this case
     of $T=0$, there are no free gluons nor energy excitations of free
     gluon medium states when the separation $R$ between the $Q$ and
     the $\bar Q$ changes, and the total free energy of the system is
     equal to the total internal energy of the system.  The real gluon
     energy degrees of freedom are absent and frozen but the virtual
     gluon degrees of freedom adjust themselves as $R$ changes.

     The question whether an adiabatic or a diabatic picture is a more
     appropriate description for the potential arises also in the NN
     and meson-meson problems at $T=0$ in lattice gauge theory.
     However aside from the question of adiabaticity, the $Q$-$\bar Q$,
     NN, and meson-meson potentials differ in their different degrees
     of freedom and the methods of calculations.  The $Q$-$\bar Q$
     potential studied here is a (two-body)-plus-(deconfined medium)
     problem, while the NN and meson-meson potentials at $T=0$ are
     (six-body)-plus-(virtual gluons) and (four-body)-plus-(virtual
     gluons) problems respectively. At $T=0$, a wave function
     treatment of the lattice gauge correlator results in the correct
     repulsive potential for the NN potential at short distances
     \cite{Ish06}, while an ``adiabatic" potential treatment without
     using the lattice wave function gives flat NN and meson-meson
     potentials \cite{Tak06}.  On the contrary, however, another
     ``adiabatic" lattice gauge meson-meson potential calculation at
     $T=0$ gives repulsive and attractive inner cores when different
     internal degrees of freedom of the light quarks are taken into
     account \cite{Det07}.  Furthermore, the lattice gauge wave
     function method of \cite{Ish06} may need additional
     justifications as questions have been raised in Appendix A of
     \cite{Det07} concerning its lattice wave function assumption.
     While the work of \cite{Ish06} appears to give a correct
     description, much work remains to be carried out to sort out the
     differences of the lattice gauge calculations of \cite{Ish06},
     \cite{Tak06}, and \cite{Det07}.  It remains another separate
     additional question how one can obtain definitive conclusions on
     the adiabaticity or diabaticity of the
     (two-body)-plus-(deconfined medium) potential at $T>T_c$ from
     these (six-body)- and (four-body)-plus-(virtual gluon) problems
     at $T=0$. As many unanswered questions remains to be resolved,
     the results of \cite{Ish06} cannot yet be used, for the present
     time at least, to draw conclusions on the adiabaticity or
     diabaticity of the $Q$-$\bar Q$ potential examined here.
     Nevertheless, the exploration of the relationship between
     adiabaticity and the shape of the relative wave function is an
     interesting subject for future investigations.

\section{An Approximate Method to Separate out the 
$Q$-$\bar Q$ Potential from $U_1$}

From the simple model of Debye screening, we observe that up to the
first order of $eV/T$, the internal energy of the medium does not
change, but up to the second order the internal energy increases with
an increasing separation between $Q$ and $\bar Q$.  This increase
arises from the fact that the thermal equilibrium attained through the
contact with a thermal bath in a grand canonical ensemble constrains
the occupation numbers of the medium particles, and this newly
re-arranged distribution leads to an increase in the number, the
entropy, and the internal energy of the medium, as a function of
increasing $R$.

Returning now to QCD lattice gauge calculations and noting its
similarities with Debye screening of Coulomb charges, we should
therefore expect that the number, the entropy, and the internal energy
of the gluon medium should likewise increase as function of increasing
$R$ between $Q$ and $\bar Q$.  Indeed, as shown for lattice gauge
calculations at a fixed temperature in Fig. 1$(a)$, there is an increase
in the entropy of the system as $R$ increases, similar to the
analogous Debye screening case shown in Fig.\ 1($b$).

Having understood the behavior of various thermodynamic quantities, we
wish to extract the $Q$-$\bar Q$ potential from lattice gauge results.
The most reliable way is to carry out additional lattice gauge
calculations to obtain $U_g^{(1)}(R)$ and $U_{g0}$.  The $Q$-$\bar Q$
potential is then the difference of $U_1(R)$ and $U_g^{(1)}(R)-
U_{g0}$, as given by Eq.\ (\ref{equqq}).  As $U_g^{(1)}(R)$ and
$U_{g0}$ in lattice gauge calculations are not yet available, we will
try to use another piece of lattice gauge data to obtain the $Q$-$\bar
Q$ potential, as least approximately.

We note that in the Debye screening case $U_{\rm medium}(R)-U_{\rm
medium}^0$ is proportional to $TS(R)$, and in the lattice gauge
calculations the quantity $TS_1(R,T)$ has been calculated.  We can look
for a similar relationship between the gluon internal energy and the
gluon entropy for the quark-gluon plasma.  If we succeed in relating
$U_g^{(1)}(R,T)-U_{g0}$ to $TS_1(R,T)$, then the $Q$-$\bar Q$
potential, $U_{Q\bar Q}^{(1)}$, can be determined from $U_1(R,T)$
by subtraction using Eq.\ (\ref{equqq}).

The subtraction can be carried out by noting that locally the
quark-gluon plasma internal energy density $\epsilon$ is related to
its pressure $p$ and entropy density $\sigma$ by the First Law of
Thermodynamics,
\begin{eqnarray}
\epsilon=T\sigma -p, 
 \end{eqnarray}
and the quark-gluon plasma pressure $p$ is also
related to the plasma energy density $\epsilon$ by the equation of
state $p(\epsilon)$ that is presumed known by another lattice gauge
calculation.  Thus, by expressing $p$ as $(3p/\epsilon) (\epsilon/3)$
with the ratio $a(T)=3p/\epsilon$ given by the known equation of
state, the plasma internal energy density $\epsilon$ is related to the
entropy density $T\sigma$ by 
\begin{eqnarray}
\epsilon=\frac{3}{3+a(T)} T\sigma .
\end{eqnarray} 
This is just
\begin{eqnarray}
\frac{dU_g^{(1)}}{dV}
=\frac{3}{3+a(T)}\frac{d}{dV}\int d{\bf r}~ T(\sigma-\sigma_0+\sigma_0),
\end{eqnarray}
where $\sigma_0$ is the entropy density in the absence of $Q$ and
$\bar Q$.  Noting that the entropy of the medium for the color-singlet
$Q$-$\bar Q$ pair is $TS_1=\int d{\bf r} T(\sigma-\sigma_0)$
and $U_{g0}$ is related to $\int d{\bf r}~T\sigma_0$, 
the above equation leads to 
\begin{eqnarray}
\frac{d[U_g^{(1)}(R,T)-U_{g0}(T)]}{dV} =\frac{3}{3+a(T)} \frac{T\,dS_1(R,T)}{dV},
\end{eqnarray} 
and the plasma internal energy integrated over the volume is given by
\begin{eqnarray}
U_g^{(1)}(R,T)-U_{g0}(T)=\frac{3}{3+a(T)}TS_1(R,T).
\end{eqnarray}
But $TS_1(R,T)$ has already been obtained as $U_1(R,T)-F_1(R,T)$.  The
plasma internal energy is therefore equal to 
\begin{eqnarray} 
U_g^{(1)}(R,T)-U_{g0}=\frac{3}{3+a(T)}[U_1(R,T)-F_1(R,T)].
\end{eqnarray}
The $Q$-$\bar Q$ potential, $U_{Q\bar Q}^{(1)}$, as determined by
subtracting the above plasma internal energy from $U_1$, is then a
linear combination of $F_1$ and $U_1$ given by \cite{Won05},
\begin{eqnarray}
\label{Uqq1}
W_1({ R},T)\equiv U_{Q\bar Q}^{(1)}({ R},T)= \frac{3}{3+a(T)} F_1({
 R},T) +\frac{a(T)}{3+a(T)} U_1({ R},T),
\end{eqnarray}
where for brevity of notation we have renamed $U_{Q\bar Q}^{(1)}({
R},T)$ as $W_1({ R},T)$ and we can define the coefficient of $F_1$,
$f_F=3/(3+a(T))$, as the $F_1$ fraction, and the coefficient of
$U_1$, $f_U=a(T)/(3+a(T))$, as the $U_1$ fraction.  The potential
$U_{Q\bar Q}^{(1)}$ is approximately $F_1$ near $T_c$ and is
approximately $3F_1/4+U_1/4$ for $T> 1.5 T_c$ \cite{Won05}.

\section{Comparison of Different $Q$-$\bar Q$ Potentials}

In the spectral function analyses, the widths of many color-singlet
heavy quarkonium states broaden suddenly at various temperatures
\cite{Asa03,Dat03,Pet05}.  In the most precise calculations for
$J/\psi$ using up to 128 time-like lattice slices, the spectrum has a
sharp peak for $0.78T_c \le T \le 1.62T_c$ and a broad structure with
no sharp peak for $1.70 T_c \le T \le 2.33T_c$ \cite{Asa03}.  The
spectral peak at the bound state has the same structure and shape at
0.78$T_c$ as it is at $1.62T_c$.  If one can infer that $J/\psi$ is
stable and bound at 0.78$T_c$, then it would be reasonable to infer
that $J/\psi$ is also bound and stable at 1.62 $T_c$.  The spectral
function at $1.70T_c$ has the same structure and shape as the spectral
function at 2.33$T_c$.  If one can infer that $J/\psi$ is unbound at
2.33$T_c$, then it would be reasonable to infer that $J/\psi$ become
already unbound at 1.70 $T_c$.  We can define the spontaneous
dissociation temperature of a quarkonium as the temperature at which
the quarkonium changes from bound to unbound and dissociates
spontaneously.  Thus, from the shape of the spectral functions, the
temperature at which the width of a $J/\psi$ quarkonium broadens
suddenly from 1.62$T_c$ to $1.70T_c$ corresponds to the $J/\psi$
spontaneous dissociation temperature.  Spontaneous dissociation
temperatures for $\chi_c$ and $\chi_b$ have been obtained in
\cite{Dat03,Pet05}.  We list the heavy quarkonium spontaneous
dissociation temperatures obtained from spectral analyses in quenched
QCD in Table I.  They can be used to test the potential models of $W_1
(\equiv U_{Q\bar Q}^{(1)})$, $F_1$, and $U_1$.

\vskip 0.4cm \centerline{Table I.  Spontaneous dissociation temperatures 
obtained from different analyses.} 
{\vskip 0.3cm\hskip 2.5cm
\begin{tabular}{|c|c|c|c|c|c|c|c|}
\hline
     & \multicolumn{4}{c|}{Quenched QCD} 
     & \multicolumn{3}{c|}{Full QCD (2 flavors)}       \\  \cline{2-7}
\hline
{\rm States     } & ~{\rm Spectral Analyses}~ 
                 &~ $W_1$
                 &  ~ $F_1$  & ~ $U_1$ 
                 &~ $W_1$ 
                 &  ~ $F_1$  & ~ $U_1$ \\

\hline
$J/\psi,\eta_c$       &   $1.62$-$1.70T_c^\dagger$ 
&   $1.62\,T_c$  
&   $1.40 \,T_c$ & $2.60 \,T_c$              
&   $1.42\,T_c$  
&   $1.21 \,T_c$ & $2.22 \,T_c$              
\\ \cline{1-8}  
$\chi_c$              & below $ 1.1T_c^{\natural}$   
&   unbound
& unbound             &  $1.18 \,T_c$            
&   $1.05 \,T_c$
&   unbound           &  $1.17 \,T_c$            
\\ \cline{1-8}
$\psi',\eta_c'$       &  
&   unbound 
&   unbound & $1.23 \,T_c$              
&   unbound  
&   unbound & $1.11 \,T_c$              
\\ \cline{1-8}  
$\Upsilon,\eta_b$     &                   
&  $ 4.1 \,T_c$    
&    $ 3.5 \,T_c$   &    $ \sim 5.0  \,T_c$                      
&  $ 3.40 \,T_c$    
&    $ 2.90 \,T_c$   &    $ 4.18 \,T_c$                           
\\ \cline{1-8}  
$\chi_b$              &  $1.15$-$1.54T_c^{\sharp}$     
&  $ 1.18 \,T_c$      
&  $ 1.10 \,T_c$      & $ 1.73 \,T_c$            
&  $ 1.22 \,T_c$      
&    $ 1.07 \,T_c$    & $ 1.61 \,T_c$            
\\ \cline{1-8}
$\Upsilon',\eta_b'$     &                   
&  $ 1.38 \,T_c$    
&    $ 1.19  \,T_c$   &    $ 2.28  \,T_c$                      
&  $ 1.18 \,T_c$    
&    $ 1.06 \,T_c$   &    $ 1.47 \,T_c$                           
\\ \cline{1-8}
                                                        \hline
\multicolumn{8}{l}
{${}^\dagger$Ref.\cite{Asa03},~~~     
 ${}^{\natural}$Ref.\cite{Dat03},~~~     
 ${}^{\sharp }$Ref.\cite{Pet05}}     
\end{tabular}
}

\vspace*{0.3cm}

To evaluate the $Q$-$\bar Q$ potentials in quenched QCD, we use the
free energy $F_1$ and the internal energy $U_1$ obtained by Kaczmarek
$et~al.$ \cite{Kac03} where $F_1$ and $U_1$ can be parametrized in
terms of a screened Coulomb potential with parameters shown in Figs. 2
and 3 of Ref.\ \cite{Won05}.  For the ratio $a(T)$ from the plasma
equation of state in Eq.\ (\ref{Uqq1}), we use the quenched equation
of state of Boyd $et~al.$ \cite{Boy96} for quenched QCD.  The quantity
$a(T)=3p/\epsilon$ and the $U_1$ and $F_1$ fractions as a function of
$T$ are shown Fig.\ 1($b$) and Fig.\ 1($c$) of Ref.\ \cite{Won05}
respectively.  These quantities allows the specification of the
$W_1\equiv U_{Q\bar Q}^{(1)}$ potential as a function of temperature.

Using quark masses $m_c=1.41$ GeV and $m_b=4.3$ GeV, we can calculate
the binding energies of heavy quarkonia and their spontaneous
dissociation temperatures using different potentials in quenched QCD.
As a function of temperature, the binding energies and wave functions
of charmonia have been presented in Fig. 6 and 7 of Ref.\ \cite{Won05}
respectively.  The bounding energies of bottomia have been presented
in Figs. 8 and 9, and the wave functions in Fig. 10 of
Ref. \cite{Won05}.  We show the root-mean-square $Q$-$\bar Q$
separation $\sqrt{\langle R^2\rangle}$ of $L=0$ charmonium calculated
with $m_c=1.41$ GeV in the $W_1(R)$ potential in quenched QCD as the
solid curve in Fig.\ 4.  As one observes, the root-mean-squared
separation $R_{\rm RMS}$ is about 1 fm for $T\sim T_c$ and it
increases to about 4 fm at $T= 1.6$$T_c$ before it becomes unbound.
The large separation is expected for systems with a weak binding, in
analogous to the halo nuclei with neutrons in weak binding observed in
nuclear physics \cite{Tan95}.  There is the question whether
charmonium with such a large separation between $Q$ and $\bar Q$ may
survive in QGP.  The dissociation cross sections for these quarkonia by
collision with gluons have been calculated and found to be a function
of the gluon collision energy, as shown in Fig.\ 13 of Ref.\
\cite{Won05}.

\begin{figure} [h]
\includegraphics[angle=0,scale=0.50]{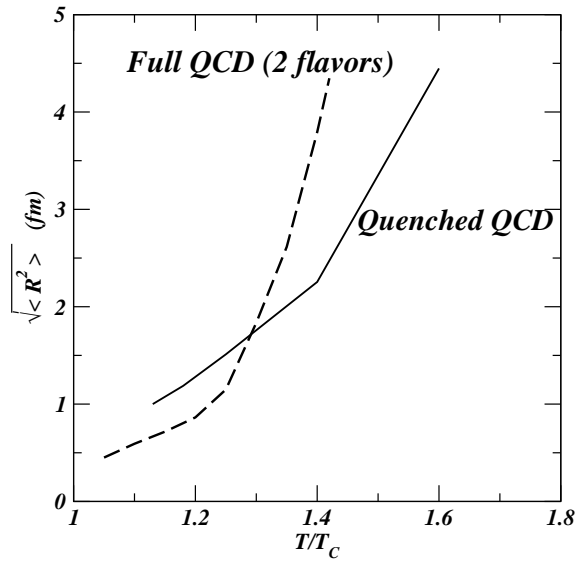}
\caption{ The root-mean-squared $Q$-$\bar Q$ separation of $L=0$
charmonium as a function of $T/T_c$ in quenched QCD (solid curve), and
in full QCD with two favors (dashed curve).  }
\end{figure}

From the binding energy of a quarkonium as a function of temperature,
one can obtain the temperature at which the quarkonium binding energy
vanishes.  This is the temperature for the spontaneous dissociation of
the quarkonium, as the quarkonium at this temperature will dissociate
spontaneously.  We list in Table I the heavy quarkonium spontaneous
dissociation temperatures calculated with the $W_1$ potential, the
$F_1$ potential, and the $U_1$ potential, in quenched QCD.

The $J/\psi$ and $\chi_b$ spontaneous dissociation temperatures
obtained with the $W_1$ potential in quenched QCD are found to be
1.62$T_c$ and $1.18T_c$ respectively.  Spectral analyses in quenched
QCD give the spontaneous dissociation temperature of 1.62-1.70$T_c$
for $J/\psi$ \cite{Asa03} and 1.15-1.54$T_c$ for $\chi_b$
\cite{Pet05}.  Thus, spontaneous dissociation temperatures obtained
with the $W_1$ potential agree with those from spectral function
analyses.  This indicates that the $W_1$ potential, defined as the
linear combination of $U_1$ and $F_1$ in Eq.\ (\ref{Uqq1}), may be the
appropriate $Q$-$\bar Q$ potential for studying the stability of heavy
quarkonia in quark-gluon plasma.

\section{$Q$-$\bar Q$ potential for full QCD with two flavors}

The interaction energy between a heavy quark and a heavy antiquark in
the color-singlet state in two-flavor full QCD was studied by
Kaczmarek and Zantow \cite{Kac05}.  In full QCD with 2 flavors, $F_1$
and $U_1$ can be represented by a color-Coulomb interaction at short
distances and a completely screened, constant, potential at large
distances as given in Ref. \cite{Won05a}, although other alternative
representations have also been presented \cite{Alb05,Dig05}.  The
transitional behavior linking the two different spatial regions can be
described by a radius parameter $r_0(T)$ and a diffuseness parameter
$d(T)$, as in the Wood-Saxon shape potential in nuclear physics,
\begin{eqnarray}
\label{scoulF}
\{F_1,U_1\}({ R},T)=
-\frac{4}{3}\frac{\alpha_s(T)}{R}f(R,T)+C(T)[1-f(R,T)],
\end{eqnarray}
where
\begin{eqnarray}
f(R,T)=\frac{1}{\exp \{(R-r_0(T))/d(T)\}+1 }.
\end{eqnarray}
In principle, it is necessary to specify only the temperature
dependence of $F_1({ R},T)$ as the internal energy $U_1({ R},T)$ can
be obtained from $F_1$ and its derivative with respect to $T$.  In
practice, as Kaczmarek and Zantow \cite{Kac05} have obtained $U_1({
R},T)$ by a careful numerical differentiation, it is convenient to
parametrize the internal energy in the above simple form for practical
calculations.

\begin{figure} [h]
\includegraphics[angle=0,scale=0.50]{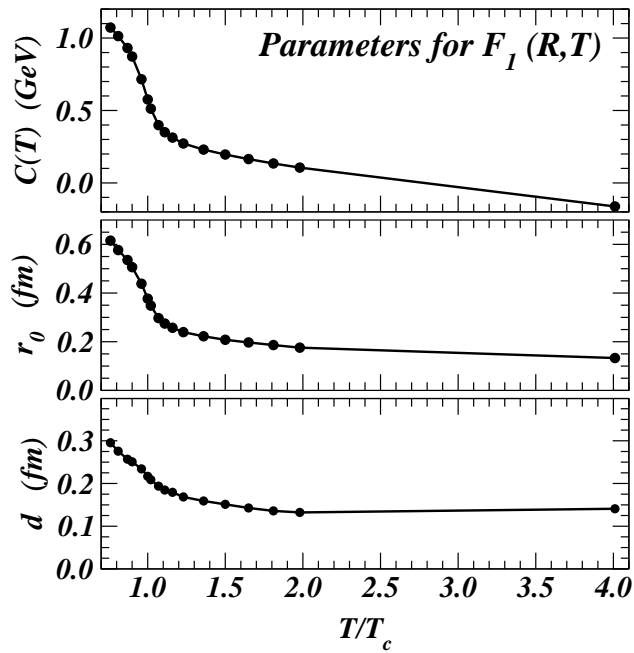}
\caption{The parameters $C$, $r_0$, and $d$ for the
color-singlet free energy $F_1({ R},T)$ in two-flavor QCD as given in
Eq.\ (\ref{scoulF}).}
\end{figure}

In searching for the coupling constant $\alpha_s$ that fits the
lattice quantities, we found that the value of $\alpha_s$ centers
around 0.3.  The fit to the lattice gauge quantities does not change
significantly whether we allow $\alpha_s$ to vary.  It is convenient
to keep the value of $\alpha_s$ to be 0.3 so that there are only three
parameters for each temperature.

\vspace*{ 0.0cm}
\hspace*{2.5cm}
\begin{figure} [h]
\centering
\includegraphics[angle=0,scale=0.45]{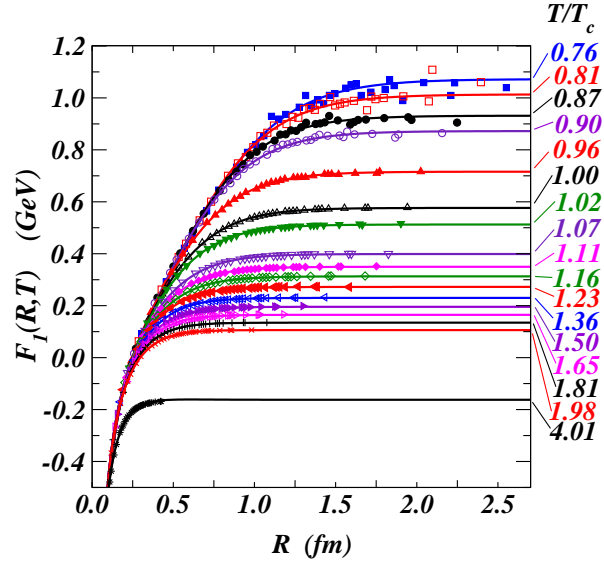}
\caption{The symbols represent the color-singlet free energy,
  $F_1(R,T)$, for two-flavor QCD \cite{Kac05}, and the curves are the
  fits using the screened potential, Eq.\ (\ref{scoulF}), with
  parameters given in Fig. 5.}
\end{figure}

For the free energy $F_1({R},T)$ in two-flavor QCD \cite{Kac05}, the
set of parameters $C$, $r_0$, and $d$ are shown in Fig.\ 5, and the
corresponding fits to $F_1$ are shown in Fig.\ 6.  For the internal
energy $U_1({R},T)$ in two-flavor QCD \cite{Kac05}, the set of
parameters $C$, $r_0$, and $d$ are shown in Fig.\ 7, and the
corresponding fits to the lattice gauge internal energy $U_1$ results
shown in Fig. 8.

\vspace*{-0.0cm}
\hspace*{1.5cm}
\begin{figure} [h]
\includegraphics[angle=0,scale=0.45]{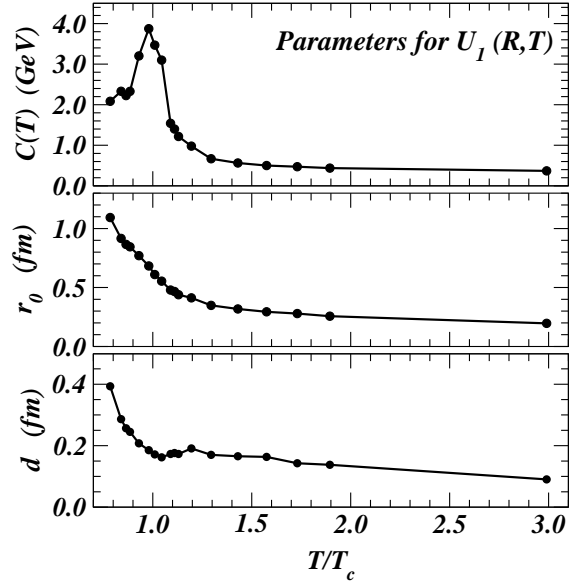}
\caption{The parameters $C$, $r_0$, and $d$ 
for the color-singlet internal energy $U_1({\bf R},T)$ as given in
Eq.\ (\ref{scoulF}).}
\end{figure}

\vspace*{-0.5cm}
\hspace*{1.5cm}
\begin{figure} [h]
\includegraphics[angle=0,scale=0.45]{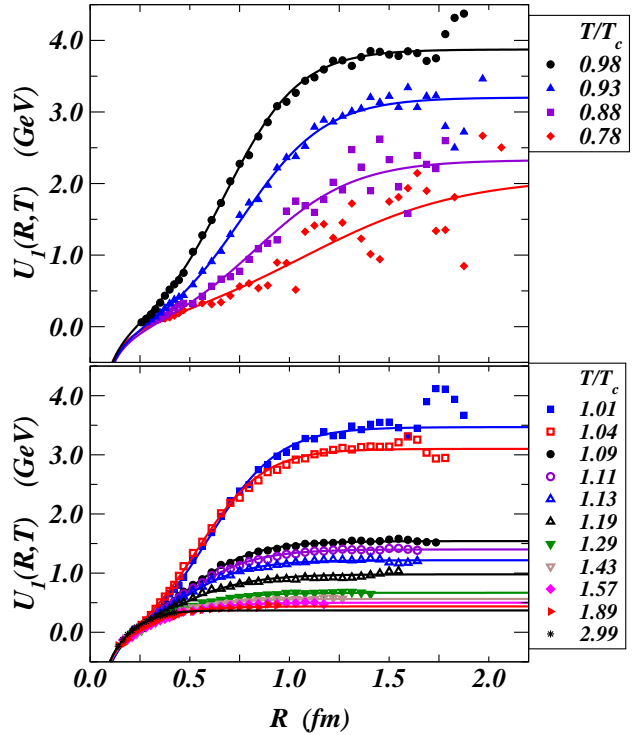}
\caption{The symbols represent $U_1(R,T)$ for two-flavor
QCD obtained by Kaczmarek $et~al.$ \cite{Kac05} and the curves are the
fits using Eq.\ (\ref{scoulF}), with parameters given in Fig. 7.  }
\end{figure}

If the thermodynamical quantity $F_1$ or $U_1$ are treated as a
potential, then the quantity $C(T)$ is an approximate measure of the
depth of the potential measured from the flat potential surface at
large distances relative to the potential well at short distances.
For the free energy $F_1$, the $C(T)$ parameter has the value of about
1 GeV at $T\sim 0.8 T_c$, and it decreases to 0.5 GeV at $T_c$.  The
free energy as a potential will has a well depth of about 0.5 GeV for
$T$ close to $T_c$ and the well depth decreases to about 0.1 GeV
at $T \sim 2 T_c$.

\hspace*{0.5cm}
\begin{figure} [h]
\includegraphics[angle=0,scale=0.45]{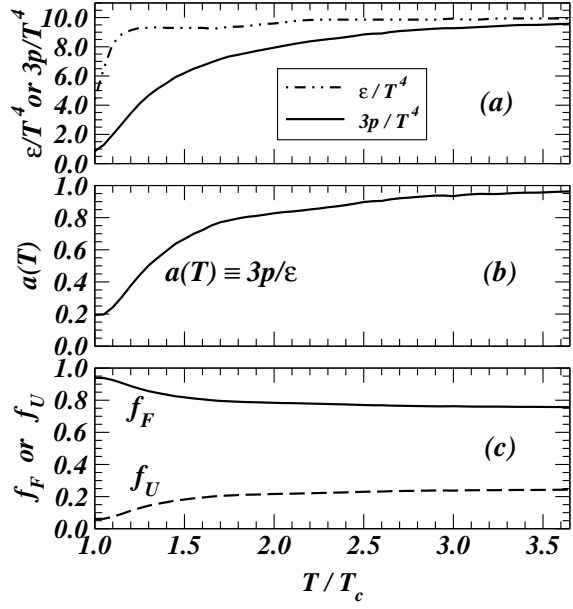}
\caption{The equation of state in full QCD with two flavors: ($a$) the
  quantity $\epsilon/T^4$ and $3 p/T^4$, ($b$) the ratio
  $a(T)=3p/\epsilon$, and ($c$) the $U_1$ and $F_1$ fractions in the
  $Q$-$\bar Q$ potential $U_{Q\bar Q}^{(1)}$. }
\end{figure}

One notes that there is a significant change in the slopes of $C(T)$
at $T\sim T_c$ for $F_1$.  As a consequence, the parameter $C(T)$ for
$U_1$ exhibits a peaks at $T\sim T_c$.  The transitional radius $r_0$
for $F_1$ decreases gradually from about 0.6 fm to about 0.15 fm and
the diffuseness parameter $d$ decreases slowly from 0.3 fm to about
0.15 fm, as temperatures decreases from 0.7$T_c$ to 4$T_c$.

For the internal energy $U_1$, the parameter $C(T)$ is quite large,
attaining the value of about 3 GeV for $T$ close to $T_c$.  This
indicates that if $U_1$ is used as a potential, the potential depth at
temperatures close to the transition temperature is of order 3 GeV,
which is a very deep potential indeed.  The parameter $C(T)$ decreases
to about 0.8 GeV when the temperature exceeds about 1.5 $T_c$.  The
transition radius $r_0$ is about 1 fm for $T$ close to 0.8 $T_c$, and
it decreases to about 0.2 fm for $T\sim 4 T_c$.  The diffuseness
parameter $d(T)$ for the internal energy decreases substantially at
temperatures below $T_c$, but maintains a relatively constant values
of 0.1 to 0.2 fm for $T$ greater than $T_c$.

The comparison in Fig. 6 and 8 shows that the free energy $F_1$ and
the internal energy $U_1$ with the set of parameters in Figs.\ 4 and
6, adequately describe the lattice-gauge data and can be used to
calculate the eigenvalues and eigenfunctions of heavy quarkonia.

In our description, the $Q$-$\bar Q$ potential is a linear combination
of $U_1$ and $F_1$ with coefficients depending on the equation of
state. The equation of state in full QCD with two flavors has been
obtained by Karsch $et~al.$ \cite{Kar00}.  We show their results of
$\epsilon/T^4$ and $3p/T^4$ in Fig.\ 9($a$).  The ratio
$a(T)=3p/\epsilon$ and the $U_1$ and $F_1$ fractions as a function of
$T$ are shown in Figs.\ 8($b$) and 8($c$) respectively.  Similar to
the case of quenched QCD, the $F_1$ fraction is close to unity near
$T_c$ and it decreases to 3/4 at large temperatures while $U_1$
fraction is nearly zero at $T_c$ and it increases to about 1/4 at high
temperatures.  These quantities, together with $F_1$ and $U_1$
specifies the $Q$-$\bar Q$ potential for bound state calculations.

\section{Heavy Quarkonia in Quark-Gluon Plasma}

In full QCD with two flavors, the transition temperature is $T_c=202$
MeV \cite{Kac05}.  To calculate the charmonium energy levels, we
employ a quark mass $m_c= 1.41$ and $m_b=4.3$ GeV.

\begin{figure}[h] 
\hspace*{0.7cm}
\includegraphics[angle=0,scale=0.50]{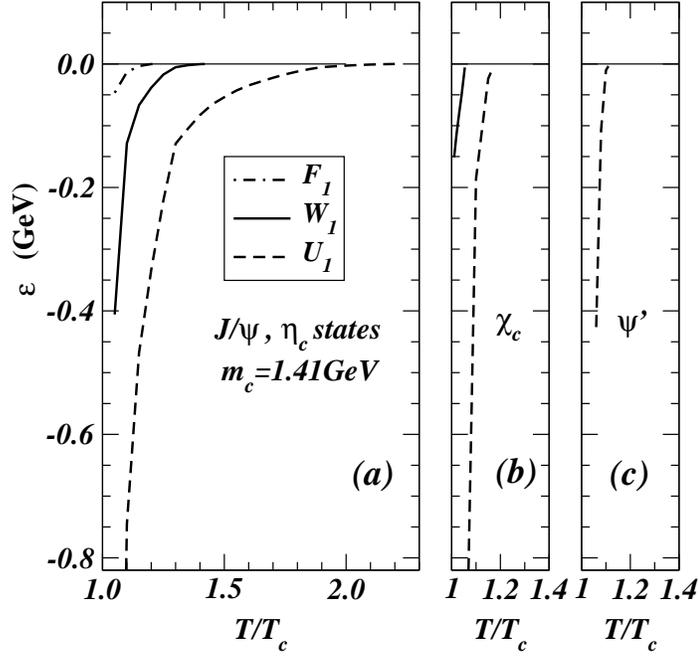} 
\caption{Energy levels of charmonium in the quark-gluon plasma as a
function of temperature calculated with the $F_1({ R},T)$,
$W_1({ R},T)$, and $U_1(R,T)$ potentials in
two-flavor QCD.  Fig. 10($a$) is for the $J/\psi$ and $\eta_c$ state,
Fig.\ 10($b$) is for the $\chi_c$ state, and Fig.\ 10($c$) is for the
$\psi'$ state.  }
\end{figure}

Energy levels of charmonium states calculated with different
potentials in full QCD with two flavors are shown in Fig.\ 10 as a
function of the temperature in units of $T_c$.  The $J/\psi$ and
$\eta_c$ states are weakly bound and they dissociate at 1.21 $T_c$ in
the $F_1$ potential, at 1.42$T_c$ in the $W_1$
potential, and at 2.22$T_c$ in the $U_1$ potential.  The $\chi_c$
state dissociates below $T_c$ in the $F_1$ potential, at 1.05$T_c$ in
the $W_1$ potential, and at 1.17$T_c$ in the $U_1$
potential.  At temperatures slightly greater than $T_c$, they are
weakly bound in the $F_1$ potential but are strongly bound in the $U_1$
potential, with a binding energy of about 0.7 GeV at 1.1$T_c$.  The
binding of the states in the $W_1$ potential lies
between these two limits.

The dissociation temperature for $J/\psi$ and $\eta_c$ in full QCD
with two flavors have been examined in the spectral function analysis
by Aarts $et~al.$ \cite{Aar06}. As the calculations have been carried
out only with a small lattice volume and a small set of statistics,
there are possible systematic uncertainties which prevented a precise
determination of the pseudocritical temperature.  Preliminary results
indicate that the $J/\psi$ state may be bound up to about $2.T_c$
\cite{Aar06}.  More definitive results will await a greater
lattice volume and larger statistics.

\begin{figure}[h]
\hspace*{0.7cm} 
\includegraphics[angle=0,scale=0.50]{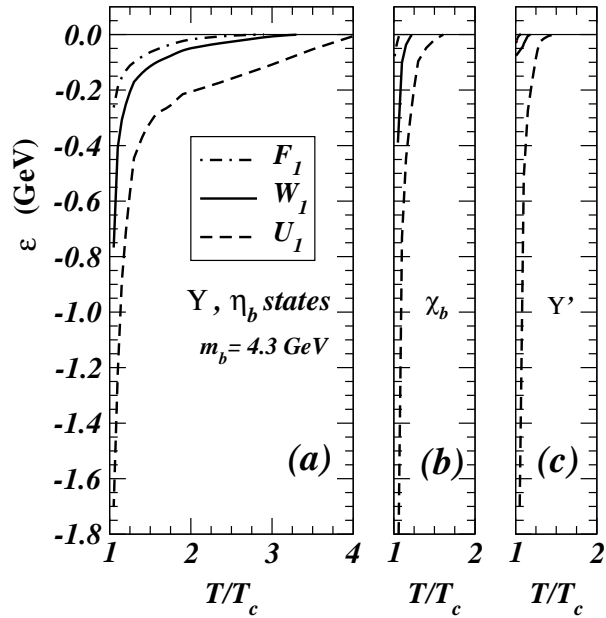}
\caption{Energy levels of bottomium in the quark-gluon plasma as a
function of the temperature calculated with the $F_1({\bf
R},T)$, $W_1({\bf R},T)$, and $U_1(R,T)$ potentials in
full QCD with two flavors.  Fig. 11($a$) is for the $\Upsilon$ and
$\eta_b$ state, Fig.\ 11($b$) is for the $\chi_b$ state, and Fig.\
11($c$) is for the $\Upsilon'$ state.  }
\end{figure}

We also carry out the analysis of bottomium 1$s$, 2$p$, and 2$s$
states.  Fig.\ 11 gives the state energies as a function of $T/T_c$
for different potentials.  The eigenenergies of $\Upsilon$ and
$\eta_b$ in two-flavor QCD are about -0.1 GeV at $T=1.1\, T_c$ in the
$F_1$ potential, and are about -1.0 GeV in the $U_1$ potential.  The
eigenenergies in the $W_1$ lie in between those of the
$F_1$ and $U_1$ potentials.  These states dissociate spontaneously at
2.9$T_c$ in the $F_1$ potential, 3.40 in the $W_1$
potential, and about 4 to 5$T_c$ in the $U_1$ potential.

In full QCD with two flavors, the $\chi_b$ state dissociates at
1.07$T_c$ in the $F_1$ potential, at 1.22 in the $W_1$
potential, and at 1.61 in the $U_1$ potential.  The $\Upsilon'$ and
$\eta_b'$ state dissociates at 1.06$T_c$ in the $F_1$ potential, at
1.18 in the $W_1$ potential, and at 1.47 in the $U_1$
potential.

In Table I we list the dissociation temperatures of different
quarkonia obtained in full QCD with two flavors.  A comparison of the
dissociation temperatures from the quenched QCD and full QCD with two
flavors show the effects of the dynamical quarks.  Dynamical quarks
increases the degree of screening, but at the same time, lowers the
phase transition temperature. They lead to a more diffused potential
with a greater screening length.  As a consequence, the binding energy
of the 1$s$ states is lowered and the dissociation temperature
decreases.  We can chooses the $W_1$ potential as the
more appropriate potential for the $Q$-$\bar Q$ pair, as it gives the
best agreement with spectral function analysis in quenched QCD.  For
this $W_1$ potential, the dissociation temperature
decreases from 1.62$T_c$ in quenched QCD to 1.42$T_c$ in full QCD with
two flavors.  The effects of the dynamical quark in full QCD leads to
a slightly weaker binding for $J/\psi$ in the plasma.  For the $\chi$
states, the effects of the the additional quark screening does not
modify the dissociation temperature substantially.  The additional
screening tends to move the centrifugal barrier for the $l=1$ state to
a smaller radial distance with a slightly higher barrier, resulting in
a very slight increase in the dissociation temperature.

\section{Quark-drip lines in Quark-Gluon Plasma}

To examine the stability of a color-singlet $Q$-$\bar Q$ pair, we
consider the quark mass $m_Q$ as a variable and evaluate the
spontaneous dissociation temperature as a function of the reduced mass
$\mu_{\rm red}=m_Q m_{\bar Q}/(m_Q + m_{\bar Q})$. The quark drip
lines calculated with the $F_1$, $W_1$, and $U_1$
potentials in quenched QCD are shown in Fig. 12.  A state is bound in
the $(T/T_c,\mu_{\rm red})$ space above a drip line and is unbound
below the drip line.  Spectral function analysis gives the spontaneous
dissociation temperature of 1.62-1.70$T_c$ for $J/\psi$
\cite{Asa03,Dat03} and $1.15$-$1.54$$T_c$ for $\chi_b$
\cite{Dat03,Pet05}.  If one takes the charm quark mass to be 1.41 GeV
and the bottom quark mass to be 4.3 GeV, the spectral function results
can be represented by the solid-circle symbols in Fig. 12.  They fall
on the drip line curves obtained with the $W_1$
potential, indicating that the $W_1$ is the appropriate
$Q$-$\bar Q$ potential to use for bound state problems.

\begin{figure} [h]
\includegraphics[angle=0,scale=0.50]{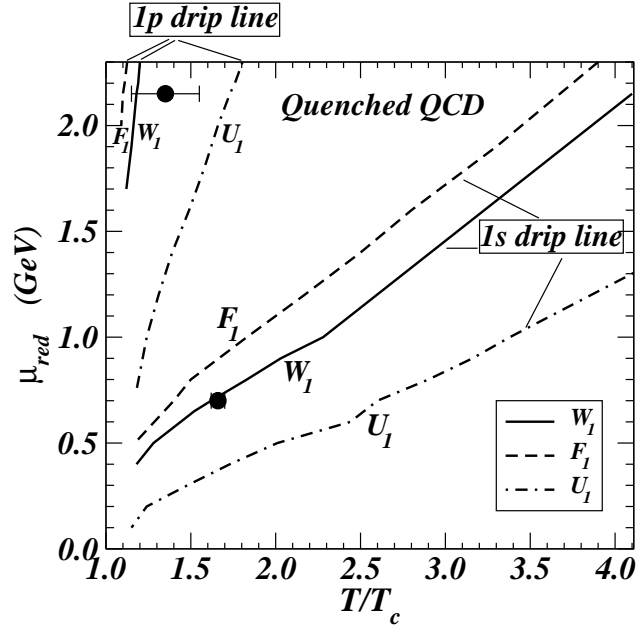}
\caption{The quark drip lines in quenched QCD calculated with the
$F_1$, $W_1$, and $U_1$ potentials.  The symbols
represent results from lattice gauge spectral function analyses.}
\end{figure}

In quenched QCD, however, the quark-gluon plasma is assumed to consist
of gluons only.  As dynamical quarks may provide additional screening,
it is necessary to consider the case with dynamical quarks.
Accordingly, we use the $F_1$, $W_1$, and $U_1$
potentials evaluated in full QCD with 2 flavors \cite{Kac05,Kar00} to
determine the drip lines in Fig.\ 13.  The drip lines for the $U_1$
potential lies lower than that of the $W_1$, which in
turn lies lower than the drip line of the $F_1$ potential.  In
comparison with quenched QCD results, the $1s$ drip line in full QCD
is shifted to lower temperatures while the $1p$ drip line in full QCD
is only slightly modified.

\begin{figure} [h]
\includegraphics[angle=0,scale=0.50]{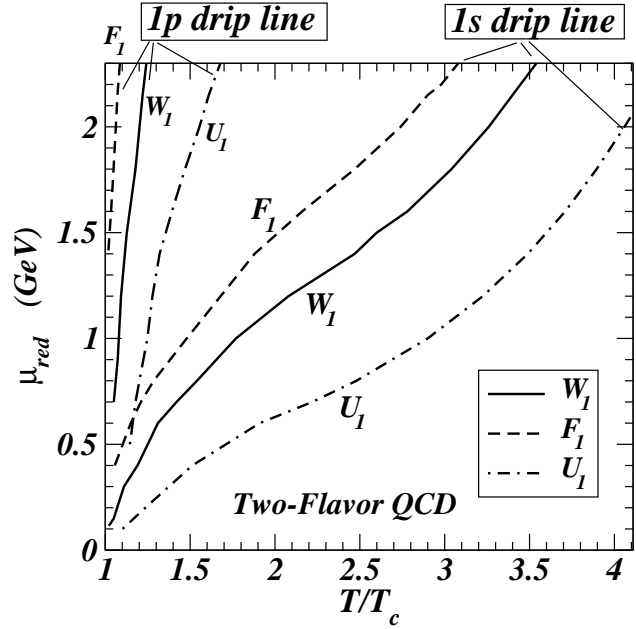}
\caption{The quark drip lines in 2-flavor QCD calculated with the
 $W_1$, $F_1$ and $U_1$ potentials.  }
\end{figure}

We shall use the results from the $W_1$ potential in
full QCD with two flavors to discuss the question of quarkonium
stability in quark-gluon plasma, as the $W_1$ potential
has been found to give results in agreement with spectral function
analyses in quenched QCD. For heavy quarkonia, results in Table I and
Fig. 13 obtained with the $W_1$ potential in full QCD
with two flavors indicate that $J/\psi$, $\chi_c$, $\Upsilon$,
$\chi_b$, and $\Upsilon'$ may be bound in the plasma up to 1.42$T_c$,
1.05$T_c$,
3.40$T_c$, 1.22$T_c$, and 1.18$T_c$ respectively.

The variation of the drip lines with the reduced mass allows us to
examine the stability of quarkonia containing quarks of various
masses.  We need to know the effective masses of different quarks in
the quark-gluon plasma.  Due to its strong interaction with other
constituents, a light quark becomes a dressed quasiparticle and
acquires a large quasiparticle mass.  In the low temperature region
where the spontaneous chiral symmetry breaking occurs with $\langle
\bar \psi \psi\rangle\ne 0$, the quasiparticle mass is $m_q\sim
[|g\langle \bar \psi \psi\rangle|$+(current quark mass)], where $g$ is
the strong coupling constant and $\langle \bar \psi \psi\rangle$ the
quark condensate \cite{Hat85,Nam61,Szc01}.  This quasiparticle mass is
the origin of the constituent-quark mass in non-relativistic
constituent quark models \cite{Hat85,Szc01,Bar92,Won02}.  In the high
temperature perturbative QCD region, the quasiparticle mass is $m_q
\sim gT/\sqrt{6}$, which is of the order of a few hundred MeV
\cite{Wel82}.

As the restoration of chiral symmetry is a second order transition,
$\langle \bar \psi \psi\rangle $ decreases gradually as the
temperature increases beyond $T_c$. The light quark quasiparticle mass
associated with $\langle \bar \psi \psi\rangle$ will likewise decrease
gradually from the constituent-quark mass value to the current-quark
mass value when the temperature increases beyond $T_c$.  This tendency
for the quasiparticle mass to decrease will be counterbalanced by the
opposite tendency for the quasiparticle `thermal mass' to increase
with increasing temperature.  As a result of these two
counterbalancing tendencies in the region of our interest,
$T_c<T<2T_c$, the effective mass of the light quarks are relatively
constant.  By examining the effects of the light quark quasiparticle
masses on the quark-gluon plasma equation of state, Levai $et~al.$
\cite{Lev98}, Szabo $et~al.$ \cite{Sza03}, and Ivanov $et~al.$
\cite{Iva05} estimate that $m_q$ is about 0.3 to 0.4 GeV at
$T_c<T<2T_c$.  As in the case of $T=0$, where light quarks with a
constituent-quark mass of about 350 MeV mimic the effects of chiral
symmetry breaking and non-relativistic constituent quark models have
been successfully used for light hadron spectroscopy
\cite{Szc01,Bar92}, so the large value of the estimated quasiparticle
mass (from 0.3 to 0.4 GeV) may allow the use of a non-relativistic
potential model as an effective tool to estimate the stability of
light quarkonia at $T_c<T<2T_c$.  It will be of interest to
investigate the relativistic effects \cite{Cra04,Cra06} in the future.

For light quark masses of 0.3 to 0.4 GeV, we can estimate from the
results in Fig. 13 for the $W_1$ potential that as a
quarkonium with light quarks has a reduced mass of 0.15-0.2 GeV, it
may be bound at temperatures below $(1.05-1.07)T_c$. An open heavy
quarkonium with a light quark and a heavy antiquark or a light
antiquark and a heavy quark have a reduced mass of about 0.3-0.4 GeV
and may be bound at temperatures below $(1.11-1.19)T_c$.

Another lattice gauge calculation gives $m_q/T=3.9\pm 0.2$ at $1.5T_c$
\cite{Pet02}, which implies that at $T$=1.5$T_c$ (or about 0.3 GeV),
the quark mass will be $\sim$1.2 GeV for $(u,d,s)$ quarks.  Such a
`light' quark quasiparticle mass appears to be quite large and may be
uncertain, as the plasma will have approximately equal abundances of
`light' and charm quarks, which is however not observed.  There may
also be difficulties in reproducing the plasma equation of state.
With this mass, a `light' quarkonium will have a reduced mass of 0.6
GeV, and the quarkonium may be bound at temperatures below
$\sim$1.31$T_c$.

In either case, the drip lines of Fig. 13 for full QCD with 2 flavors
obtained with the $W_1$ potential do not support bound
$Q\bar Q$ states with light quarks beyond 1.5$T_c$. A recent study of
baryon-strangeness correlations suggests that the quark-gluon plasma
contains essentially no bound $Q\bar Q$ component at 1.5$T_c$
\cite{Koc05}.

\section{Conclusions and Discussions}

The degree to which the constituents of a quark-gluon plasma (QGP) can
combine to form composite entities is an important property of the
plasma.  To study the composite nature of the plasma, we need to
examine the stability of quarkonium in quark-gluon plasma which
depends on the $Q$-$\bar Q$ potential.  We seek to extract the
$Q$-$\bar Q$ potential from thermodynamical quantities obtained in
lattice gauge calculations.  For such a purpose, we need the
relationship between the $Q$-$\bar Q$ potential and the internal
energy obtained in lattice gauge calculations.  Such a relationship
was derived previously in Ref.\ \cite{Won05}.  We would like to gain
additional support by examining whether a similar relationship exists
between the $Q$-$\bar Q$ potential and the internal energy in an
analogous, but not identical, case of Debye screening.

We find that in adiabatic motion of $Q$ and $\bar Q$ under Debye
screening, (1) the potential for the $Q$ and $\bar Q$ in the Schr\"
odinger equation contains the interactions that act on $Q$ and $\bar
Q$, (2) this $Q$-$\bar Q$ potential under Debye screening is only part
of the total internal energy of the system, (3) the other part of the
internal energy is the internal energy of the medium particles, and
(4) many thermodynamical quantities such as the number, the entropy,
and the internal energy of the medium particles increases with the
separation between $Q$ and $\bar Q$ in the grand canonical ensemble.
Therefore, to obtain the Debye screening potential between two static
charges, it is necessary to subtract out the internal energy of the
medium particles from the total internal energy.  These results
supports a similar conclusion reached earlier in the analogous lattice
gauge theory \cite{Won05}.

We are thus led to obtain the $Q$-$\bar Q$ potential in the
quark-gluon plasma by subtracting out the internal energy of the
medium particles from the total internal energy in the grand canonical
ensemble.  We proposed a method to subtracting out the internal energy
of the medium by making use of the equation of state of the
quark-gluon plasma obtained in an independent lattice gauge
calculation \cite{Won05}.  The potential can then be represented as a
linear combination of $U_1$ and $F_1$, with coefficients depending on
the quark-gluon plasma equation of state.  The proposed potential in
the quenched approximation is found to give dissociation temperatures
that agree with those from spectral function analyses.  It can be
generalized to the case of full QCD to discuss quarkonium states in
the plasma.

The knowledge of the single-particle states using potentials extracted
from lattice gauge calculations in full QCD can then be used to
examine the limit of stability of both heavy and light quarkonia and
to determine the location of the quark drip lines.

The quark drip lines allows one to ascertain the degree of stability
of heavy and light quarkonia when the masses of the quarks are known.
$J/\psi$, $\chi_c$, $\Upsilon$, $\chi_b$, and $\Upsilon'$ are found to
be stable in the plasma and dissociate at different temperatures. The
characteristics of the quark drip lines severely limit the region of
possible quarkonium states with light quarks to temperatures close to
the phase transition temperature.  Various estimates give a light
quark mass of about 0.3-0.4 GeV \cite {Lev98,Sza03,Iva05}, which is
not very different from the constituent quark masses in
non-relativistic quark models of hadrons.  Bound quarkonia with light
quarks may exist very near the phase transition temperature if their
effective quark mass is of the order of 300-400 MeV and higher.

\vspace*{0.3cm} The author thanks Drs. H. Crater and Su-Houng Lee for
helpful discussions.  This research was supported in part by the
Division of Nuclear Physics, U.S. Department of Energy, under Contract
No. DE-AC05-00OR22725, managed by UT-Battle, LLC and by the National
Science Foundation under contract NSF-Phy-0244786 at the University of
Tennessee.

\end{document}